\documentclass[apj]{emulateapj}

\usepackage{graphicx,booktabs,float,color}
\shorttitle{Grain Growth in the Disks of CY Tau and DoAr 25}
\shortauthors{P\'erez et al.}

\newcommand{\cytau}{CY\nobreak\hspace{.16667em plus .08333em}Tau}
\newcommand{\doardisk}{DoAr\nobreak\hspace{.16667em plus .08333em}25}

\begin{document}

\title{Grain Growth in the Circumstellar Disks of the Young Stars CY Tau and DoAr 25}

\author{Laura M. P\'erez\altaffilmark{1,2},
Claire J. Chandler\altaffilmark{1},
Andrea Isella\altaffilmark{3}, 
John M. Carpenter\altaffilmark{4}, 
Sean M. Andrews\altaffilmark{5},
Nuria Calvet\altaffilmark{6},
Stuartt A. Corder\altaffilmark{7}, 
Adam T. Deller\altaffilmark{8}, 
Cornelis P. Dullemond\altaffilmark{9}, 
Jane S. Greaves\altaffilmark{10},
Robert J. Harris\altaffilmark{11},
Thomas Henning\altaffilmark{12},
Woojin Kwon\altaffilmark{13},
Joseph Lazio\altaffilmark{14},
Hendrik Linz\altaffilmark{12},
Lee G. Mundy\altaffilmark{15}, 
Luca Ricci\altaffilmark{5},
Anneila I. Sargent\altaffilmark{4},
Shaye Storm\altaffilmark{15}, 
Marco Tazzari\altaffilmark{16},
Leonardo Testi\altaffilmark{16,17},
David J. Wilner\altaffilmark{5}
}
\altaffiltext{1}{National Radio Astronomy Observatory, P.O. Box O, Socorro, NM 87801, USA}
\altaffiltext{2}{Jansky Fellow of the National Radio Astronomy Observatory}
\altaffiltext{3}{Rice University, 6100 Main Street, Houston, TX 77005, USA}
\altaffiltext{4}{California Institute of Technology, 1200 East California Blvd, Pasadena, CA 91125, USA}
\altaffiltext{5}{Harvard-Smithsonian Center for Astrophysics, 60 Garden Street, Cambridge, MA 02138, USA}
\altaffiltext{6}{University of Michigan, 830 Dennison Building, 500 Church Street, Ann Arbor, MI 48109, USA}

\altaffiltext{7}{Joint ALMA Observatory, Av. Alonso de C\'ordova 3107, Vitacura, Santiago, Chile}
\altaffiltext{8}{The Netherlands Institute for Radio Astronomy (ASTRON), 7990-AA Dwingeloo, The Netherlands}
\altaffiltext{9}{Heidelberg University, Center for Astronomy, Albert Ueberle Str 2, Heidelberg, Germany}
\altaffiltext{10}{University of St. Andrews, Physics and Astronomy, North Haugh, St Andrews KY16 9SS}
\altaffiltext{11}{University of Illinois, 1002 West Green St., Urbana, IL 61801, USA}
\altaffiltext{12}{Max-Planck-Institut f\"{u}r Astronomie, K\"{o}nigstuhl 17, 69117 Heidelberg, Germany}
\altaffiltext{13}{Korea Astronomy and Space Science Institute, 776 Daedeok-daero, Yuseong-gu, Daejeon 34055, Korea}
\altaffiltext{14}{Jet Propulsion Laboratory, California Institute of Technology, 4800 Oak Grove Dr, Pasadena, CA  91106}
\altaffiltext{15}{University of Maryland, College Park, MD 20742, USA}
\altaffiltext{16}{European Southern Observatory, Karl Schwarzschild str. 2, 85748 Garching, Germany}
\altaffiltext{17}{INAF-Osservatorio Astrofisico di Arcetri, Largo E. Fermi 5, 50125 Firenze, Italy}

\begin{abstract}
\noindent 
We present new results from the Disks@EVLA program for two young stars: \cytau\ and \doardisk. We trace continuum emission arising from their circusmtellar disks from spatially resolved observations, down to tens of AU scales, at $\lambda=0.9, 2.8, 8.0, 9.8$~mm for \doardisk\ and at $\lambda=1.3, 2.8, 7.1$~mm for \cytau. Additionally, we constrain the amount of emission whose origin is different from thermal dust emission from 5~cm observations. Directly from interferometric data, we find that observations at 7~mm and 1~cm trace emission from a compact disk while millimeter-wave observations trace an extended disk structure. From a physical disk model, where we characterize the disk structure of \cytau\ and \doardisk\ at wavelengths shorter than 5~cm, we find that (1) dust continuum emission is optically thin at the observed wavelengths and over the spatial scales studied, (2) a constant value of the dust opacity is not warranted by our observations, and (3) a high-significance radial gradient of the dust opacity spectral index, $\beta$, is consistent with the observed dust emission in both disks, with low-$\beta$ in the inner disk and high-$\beta$ in the outer disk. Assuming that changes in dust properties arise solely due to changes in the maximum particle size ($a_{max}$), we constrain radial variations of $a_{max}$ in both disks, from cm-sized particles in the inner disk ($R<40$~AU) to millimeter sizes in the outer disk ($R>80$~AU). These observational constraints agree with theoretical predictions of the radial-drift barrier, however, fragmentation of dust grains could explain our $a_{max}(R)$ constraints if these disks have lower turbulence and/or if dust can survive high-velocity collisions.

\end{abstract}


\section{Introduction}
The process of planet formation requires that small dust grains in the interstellar medium (ISM) undergo a dramatic transformation, growing by more than 12 orders of magnitude in order to form terrestrial planets and the cores of gas and ice giants. 
As a very first step, these $\mu$m-sized ISM dust grains must increase their size and become macroscopic \citep{2014Testi}, a process that alters the optical properties of the dust grains considerably: as grains reach millimeter or larger sizes, the absolute value of the dust opacity, $\kappa_{\lambda}$, decreases, and at the same time the power-law spectral index of the dust opacity, $\beta$ (where $\kappa_{\lambda}\propto\lambda^{-\beta}$), becomes smaller \citep[e.g.][]{1993Miyake,1996Henning,2006Draine}. 
Since thermal dust emission at millimeter and centimeter wavelengths is (mostly) in the optically thin regime, the observed dust continuum emission will trace the bulk of the disk mass modulated by the dust opacity: $S_{\nu}\propto\kappa_{\lambda}\times \Sigma \times B_{\nu}(T)$, where $\Sigma$ is the dust mass surface density and $B_{\nu}(T)$ is the Planck function evaluated at the dust temperature $T$. 
For optically thin dust, warm enough to be in the Rayleigh-Jeans (R-J) regime ($h\nu \ll kT$), it follows that $S_{\nu}\propto\lambda^{-(\beta+2)}$.
Thus, direct measurements of the dust emission spectrum (i.e., the spectral energy distribution, SED, at long wavelengths) 
can be used to derive the value of $\beta$, a method extensively employed in the literature \citep{1991Beckwith,1995P&SS...43.1333H,2001Testi,2002Calvet,2003Testi,2004Natta,2005Wilner,2005Andrews,2006Rodmann,2007Andrews,2009Lommen,2010Ricci_a,2010Ricci_b,2011Ricci,2012Ubach}.

Disk-integrated measurements, from sub-mm to cm wavelengths, have shown that in most protoplanetary disks the value of $\beta$ tends to be lower than in the ISM \citep[where $\beta_{\rm ISM}\sim1.5-2.0$, consistent with the presence of $\mu$m-sized dust grains or smaller in the interstellar medium, ][]{2001Li_Draine}.

Possible explanations for the low values of $\beta$ include the presence of regions of high optical depth, which will drive the dust emission spectrum to be close to $S_{\nu}\propto\lambda^{-2}$, and thus a low $\beta$ would be inferred (incorrectly). 
However, \citet{2012Ricci} has shown this would be only plausible for the brightest and most massive disks.
Different composition and/or porosity could affect the optical properties of dust grains \citep[e.g., ][]{1997Henning,2003Semenov,2005Boudet,2014Kataoka}, but for a sensible set of dust properties the most significant influence on the opacity spectral index arises from an increase in grain size \citep{2006Draine}.
Thus, the difference in the dust opacity spectral index of disks, as compared to the ISM, implies that growth of at least 4 orders of magnitude in size has taken place inside circumstellar disks \citep[for a review see: ][]{2007Natta,2014Testi}.
But since most of these measurements are of the global disk properties, they provide limited knowledge of the evolution of the dust properties as a function of the distance to the central star (hereafter referred to as radius).

Theoretical analyses predict that the average dust grain size will change with radius, as dust grows and fragments while it is transported throughout the disk 
\citep[][]{2005DullemondDominik,2010Birnstiel}.
A fundamental limiting factor to the largest grain size is that of the radial drift problem. 
While solids move at Keplerian velocities in the disk, the gaseous component moves at slightly sub-Keplerian speed due to gas pressure support. 
This velocity difference induces a drag in the large particles, which end up losing angular momentum and drifting radially to smaller and smaller orbits, until they are accreted onto the star and lost from the solid population \citep{1977Weidenschilling}. 
This barrier will limit the maximum grain size allowed in a disk, with mm and cm-sized particles at tens of AU from the star drifting inwards in timescale shorter than the disk lifetime \citep{2010Youdin}. 
Thus, measuring changes in the dust properties as a function of radius is essential to investigate the effects of this barrier in the grain size distribution of disks.

Only recently have measurements reached the required sensitivity and a sufficient lever arm in wavelength to characterize radial variations of the dust properties \citep{2010Isella,2011Guilloteau,2011Banzatti,2012Perez,2013Trotta,2014Andrews,2014Menu}.
With wavelength coverage that spanned an order of magnitude, from sub-mm to cm wavelengths, observational constraints in dust opacity were obtained for the disk around the young star AS~209 \citep{2012Perez}.
In this study, a gradient in the value of $\beta$ was found from $\beta<0.5$ at $\sim$20~AU to $\beta>1.5$ at more than $\sim$80~AU, while a constant value of the dust opacity throughout the disk failed to reproduce these observations.
Furthermore, these results seem to agree with a population of dust grains limited by radial drift.
Here we explore further this issue, by studying how the dust properties vary with radius in the protoplanetary disks that surround the young stars \cytau, located in the Taurus star forming region at a distance of $\sim140$~pc \citep{2007ApJ...671.1813T}, and \doardisk, located in the L1688 dark cloud at a distance of $\sim125$~pc \citep{2008ApJ...675L..29L,2008AN....329...10M}.
\doardisk\footnote{also known as GY92~17, WSB~29, and WLY~1-34} and \cytau\ are pre-main sequence stars of K5 and M1 spectral type, respectively. 
Both stars are quite young: \cytau\ is $2-3$ Myr old \citep{2007A&A...473L..21B,2014A&A...567A.117G}, while \doardisk\ is about 4 Myr old \citep{2009Andrews}, and both present a significant emission excess over the stellar photosphere, from near-infrared to millimeter wavelengths \citep{2009A&A...507..327O,2010ApJS..188...75M}. \doardisk\ has been imaged at sub-millimeter wavelengths with the Submillimeter Array \citep{2008Andrews,2009Andrews}, and \cytau\ was imaged at the 1.3 and 2.8~mm bands with the Plateau de Bure Interferometer \citep{2011Guilloteau}.

In this paper, we triple the number of circumstellar disks for which our multiwavelength analysis has been used to measure any radial changes in the dust properties, particularly in the dust opacity.
The paper is structured as follows: first, we describe the observations and data calibration procedures in Section~\ref{ObsData}, and we present observational results from these data in Section~\ref{ObservationalResults}. In Section~\ref{modeling_emission} we discuss the model of the disk emission employed in this paper, followed by Section~\ref{results_modeling} where observational constraints derived from this modeling are presented. Finally, in Section~\ref{discussion} we discuss our results in the context of grain growth, comparing these observational constraints with the radial drift and fragmentation barriers for growth, as well as with other disks previously studied. Our findings and conclusions are presented in Section~\ref{conclusions}.

\section{Observations}
\label{ObsData}
We obtained multi-wavelength observations of the continuum emission from \doardisk\ and \cytau\ with three interferometers: the Combined Array for Research in Millimeter-wave Astronomy (CARMA), the Sub-Millimeter Array (SMA), and the Karl G. Jansky Very Large Array (VLA) as part of the Disks@EVLA collaboration. \cytau\ was observed at four different wavelengths: 1.3~mm, 2.8~mm, 7.1 mm, and 5.0 cm, with the highest angular resolution being $0.07''$ (9.8~AU) at 7.1~mm.
\doardisk\ was observed at five different wavelengths: 0.9~mm, 2.8~mm, 8.0~mm, 9.8~mm, and 5.0~cm, with the highest angular resolution being $0.10''$ (12.5~AU) at 8.0~mm. A summary of these observations and our data calibration procedure is presented below.

\begin{deluxetable}{l c l c}
\tabletypesize{\footnotesize}
\tablewidth{0pc}
\tablecaption{Observing log for CARMA observations}
\tablecolumns{4}
\tablehead{
\colhead{Target} & \colhead{UT Date} & \colhead{Configuration\tablenotemark{a}} & \colhead{Phase Calibrator}
}
\startdata
\cutinhead{$\lambda=1.3$ mm}
CY Tau	& 2007-Nov-07 &	C, 0.35 km & J0530+1331	\\
		& 2009-Jan-04 &	B, 1.0 km & 3C 111	    \\
        & 2009-Jan-05 &	B, 1.0 km & 3C 111		\\
		& 2010-Sep-16 &	D, 0.25 km & J0336+3218    \\
		& 2010-Sep-21 &	D, 0.25 km & 3C 111, J0336+3218 \\
        & 2011-Dec-12 & A, 2.0 km & J0336+3218    \\

\cutinhead{$\lambda=2.8$ mm}
CY Tau	& 2008-Dec-11 &	B, 1.0 km   & 3C 111		    \\
        & 2010-Feb-15 & A, 2.0 km	& 3C 111,J0336+3218 \\
        & 2011-Dec-30 &	B, 1.0 km   & 3C 111		    \\
		& 2012-Feb-08 &	C, 0.35 km   & 3C 111		    \\
		& 2012-Feb-19 &	C, 0.35 km   & 3C 111		    \\

\hline

DoAr 25	& 2010-Jan-06 & B, 1.0 km	& J1625-2527         \\
		& 2010-Jan-09 & B, 1.0 km	& J1625-2527         \\
        & 2010-Nov-29 & A, 2.0 km	& J1625-2527         \\
		& 2010-Nov-30 & A, 2.0 km	& J1625-2527         \\
		& 2010-Dec-03 & A, 2.0 km	& J1625-2527         \\
		& 2010-Dec-12 & A, 2.0 km	& J1625-2527         \\
		& 2011-Apr-25 & C, 0.35 km	& J1625-2527         \\
		& 2012-Jan-19 & C, 0.35 km	& J1625-2527         \\
\enddata
\tablenotetext{a} {Configuration name followed by the maximum baseline of each configuration.}
\label{CARMA_table}
\end{deluxetable}

\begin{deluxetable}{l c l c}
\tabletypesize{\footnotesize}
\tablewidth{0pc}
\tablecaption{Observing log for VLA observations}
\tablecolumns{4}
\tablehead{
\colhead{Target} &  \colhead{UT Date} & \colhead{Configuration\tablenotemark{a}} & \colhead{Phase Calibrator}}
\startdata
\cutinhead{$\lambda=7.1$ mm (Q band)}
CY Tau	& 2010-Nov-13	& C, 3.4 km		& J0403+2600 \\
		& 2011-Apr-05	& B, 11.1 km	& J0403+2600 \\
		& 2012-Oct-22	& A, 36.4 km		& J0403+2600 \\
		& 2012-Oct-26	& A, 36.4 km		& J0403+2600 \\
		& 2012-Oct-27	& A, 36.4 km		& J0403+2600 \\
		& 2012-Oct-28	& A, 36.4 km		& J0403+2600 \\

\cutinhead{$\lambda=8.0$ and 9.8 mm (Ka band)}
DoAr 25	& 2011-Jan-23	& CnB, 11.1 km	& J1625-2527 \\
		& 2011-May-28	& BnA, 36.4 km	& J1625-2527 \\
		& 2011-Jun-15	& A, 36.4 km		& J1625-2527 \\

\cutinhead{$\lambda=5.0$ cm (C band)}
CY Tau	& 2011-Jul-23	& A, 36.4 km		& J0403+2600 \\
DoAr 25	& 2011-Jul-14	& A, 36.4 km		& J1625-2527 \\
	
\enddata
\tablenotetext{a} {Configuration name followed by the maximum baseline of each configuration.}
\label{VLA_table}
\end{deluxetable}

\subsection{CARMA Observations}

\cytau\ and \doardisk\ observations at 2.8~mm (107~GHz), and \cytau\ observations at 1.3~mm (230~GHz) were obtained with CARMA in the A, B, and C configurations, providing baseline lengths between 30--2000~m which cover spatial scales from $20''$ down to $0.3''$ at 2.8~mm, and between $9''$ down to $0.13''$ at 1.3~mm. A complete observing log of these observations can be found in Table \ref{CARMA_table}.
Double-sideband single-polarization receivers were tuned to a frequency of 107~GHz in the case of 2.8~mm observations, and 230~GHz in the case of 1.3~mm observations. To optimize the continuum sensitivity, we configured all spectral windows in the correlator to the maximum possible bandwidth: 468.75-MHz-wide windows of 15 channels each, which provided 2.8 GHz of bandwidth for observations before the correlator upgrade in December 2009, and 487.5-MHz-wide windows of 39 channels each, which provided 7.8 GHz of total bandwidth after the upgrade. 

The observing sequence interleaved observations of a complex gain calibrator (generally 3 minutes in length) with the science target. In C configuration the calibrator-target cycle time was 12--15 minutes, while in A and B configuration it was 5--10 minutes in order to track better the tropospheric phase fluctuations.  This observing sequence repeated throughout the track, usually from target rise to set.  In addition, a strong calibrator was observed to measure the complex bandpass, along with either a planet (Uranus, Mars, or Neptune) or a secondary flux density calibrator (either 3C273 or 3C84) that is monitored by the observatory. The estimated fractional uncertainty in the absolute flux density scale is $\sim15$\%.

The CARMA data were calibrated using the Multichannel Image Reconstruction, Image Analysis and Display (MIRIAD) software package \citep{1995Sault}. Each observation was calibrated separately. Malfunctioning antennas, receivers, and/or spectral windows were flagged, and updated antenna position corrections and line-length system corrections were applied. Observations of \doardisk\ in the most extended (A and B) configurations used the CARMA Paired Antenna Calibration System (C-PACS) to monitor the tropospheric delay fluctuations on eight of the outermost stations of the array, using adjacent 3.5~m antennas equipped with 1~cm receivers.  Phase corrections derived from the C-PACS system are applied during post-processing. A complete description of C-PACS is presented by \citet{2010ApJ...724..493P} and \citet{2014arXiv1410.5560Z}.

\subsection{VLA Observations}

\subsubsection{High Frequency}

\cytau\ observations at 7.1~mm (42~GHz, Q-band) were made during A, B and C configurations of the VLA, while \doardisk\ observations at 8.0 and 9.8~mm (30.5~Ghz and 37.5~GHz, Ka-band) were made during A, BnA, and CnB  configurations,\footnote{The hybrid BnA and CnB configurations comprise antennas on the east and west arms of the VLA moved to their locations for the more compact configuration, while the antennas on the north arm are moved to their locations for the more extended configuration, so that in projection, for southern sources, the synthesized beam is circularized.} providing baseline lengths for both targets between 60~m to 36~km. These baseline lengths correspond to spatial scale coverage from about $24''$ down to $0.04''$ at 7.1~mm, and from about $34''$ down to $0.06''$ at 9.8~mm.
These observations, summarized in Table \ref{VLA_table}, used dual-polarization receivers and two independently tunable basebands, where each baseband was configured to eight 128~MHz spectral windows of 64 channels, to provide the maximum continuum bandwidth per baseband (1~GHz).
For the Q-band observations, the two basebands were centered at 7.1~mm (42.5~GHz) and 7.2~mm (41.5~GHz) providing 2~GHz total bandwidth at 7.14~mm, while for the Ka-band observations, the two 1~GHz basebands were centered at 8.0~mm (37.5~GHz) and 9.8~mm (30.5~GHz).

\begin{figure*}
\begin{center} 
\includegraphics[scale=0.57]{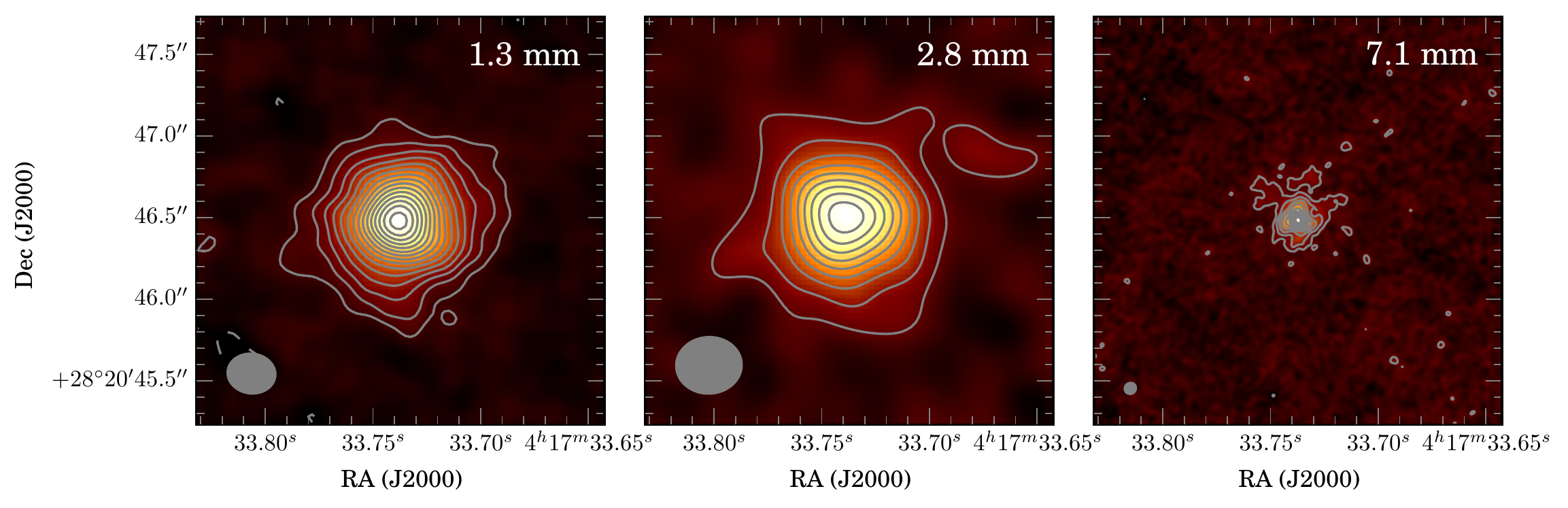}
\caption{Aperture synthesis images of the continuum emission towards the young star \cytau, observed at wavelengths of 1.3, 2.8, and 7.1~mm. 
Each panel encompasses a $2.5''\times2.5''$ region (350~AU in size at the adopted distance), with contours drawn at $3\sigma$ intervals, where $\sigma$ is the RMS noise level in each map (see Table \ref{table_image_cytau}).}
\label{maps_cytau}
\end{center}
\end{figure*}

\subsubsection{Low Frequency}

To distinguish the amount of free-free or non-thermal contamination present at millimeter and centimeter wavelengths, \cytau\ and \doardisk\ were observed at 5.0 cm (6.0 GHz) in C-band.
These low-frequency VLA observations were obtained during A configuration, with dual-polarization receivers and two independently tunable basebands, where each baseband was configured as for the high frequency observations with a total bandwidth per baseband of 1~GHz. The two basebands were centered at 6.2~cm (4.8~GHz) and 4.1~cm (7.3~GHz), in order to be able to infer the spectral slope of the emission at centimeter wavelengths.

\subsubsection{VLA data acquisition and calibration}

The observing sequence interleaved observations of a complex gain calibrator (45~sec to a few minutes in length) with science target observations, whose duration depended on the array configuration and observing frequency: target-calibrator cycle times in extended configurations were 1.5--5 minutes at high frequencies, 5--10 minutes in compact configurations. The target-calibrator cycle time at C-band in A-configuration was 10--15 minutes.  A strong calibrator was used to determine the bandpass shape. The absolute flux density calibration was determined from observations of a primary flux density calibrator (3C147 for \cytau, 3C286 for \doardisk). Models for the primary flux density calibrators are provided by the observatory.  The estimated uncertainty in the absolute flux density scale is 5\% at C-band, and 10\% at Ka- and Q-bands.

The VLA data calibration was performed using the CASA software package and a modified version of the VLA calibration pipeline\footnote{https://science.nrao.edu/facilities/vla/data-processing/pipeline}. As with the CARMA observations, each observation was calibrated separately. Malfunctioning antennas, receivers, and/or spectral windows were flagged, as well as noisy channels at the edge of each spectral window. The first few seconds at the beginning of each target observation were also flagged, along with any radio frequency interference (especially an issue with the C-band data). Times of poor phase coherence, as measured on the gain calibrator scans, were also removed from the data. Because the data were acquired over an extended time period, corrections for source proper motion or other systematic position offsets were applied prior to combining.  All VLA data presented here were obtained as part of the Disks@EVLA project (AC982).

\subsection{SMA Observations}

\doardisk\ observations at 0.9~mm (345~GHz) were obtained with the SMA between 2005 May and 2007 June; these observations have already been presented by \cite{2008Andrews,2009Andrews}. Three different array configurations (C, E, and V) were used, providing baseline lengths between 8-500~m, corresponding to spatial scales from about $22''$ down to $0.3''$. Double-sideband receivers were tuned to a local oscillator frequency of 340.755~GHz. Each sideband was configured to have 24 partially overlapping 104~MHz chunks, for a total continuum bandwidth of 4~GHz. The observing sequence interleaved observations of a complex gain calibrator with science target observations twice as long. The source-calibrator cycle was 8 minutes for the extended configuration, and $\sim15$--20 minutes for compact configurations. Bandpass and flux density calibrators were selected from different planets and satellites (Uranus, Jupiter, Saturn, Titan, Callisto), as well as strong quasars (3C454.3, 3C279), depending on their availability and array configuration. The estimated uncertainty in the absolute flux density scale is $\sim 10\%$. The data were flagged and calibrated with the IDL-based MIR software package.

\subsection{Averaging of interferometric data}

For each telescope we verified that the calibration from multiple observations obtained in different array configurations matched, to within the uncertainty in the absolute flux scale, by comparing the calibrated visibilities where they overlap in {\it uv}-space.  
The observations at 0.9, 1.3, 2.8, 7.1, 8.0, and 9.8~mm were then averaged over the observing bandwidth, as follows.
In the case of the 0.9, 1.3, 2.8 and 7~mm data, visibilities encompass a narrow range of frequencies over the entire bandwidth: $\Delta\lambda/\lambda \sim 0.01$ at 0.9~mm, $\Delta\lambda/\lambda \sim 0.04$ at 1.3~mm, $\Delta\lambda/\lambda \sim 0.08$ at 2.8~mm, and $\Delta\lambda/\lambda \sim 0.05$ at 7.1~mm. These data were therefore averaged into a single wideband channel.
In the case of Ka-band VLA observations, the two 1-GHz wide basebands are separated by $\sim7$~GHz, so these data were averaged into two wideband channels: one centered at 30.5 GHz ($\lambda = 9.8$~mm), the other centered at 37.5~GHz ($\lambda = 8.0$~mm).
Calibrated observations from different array configurations were combined to generate a single visibility file for each wavelength analyzed, in order to compare these interferometric observations with physical disk models.

\section{Observational Results}
\label{ObservationalResults}
\subsection{CY Tau}
\label{cytau_obs_res}

\begin{figure*}
\begin{center} 
\includegraphics[scale=0.57]{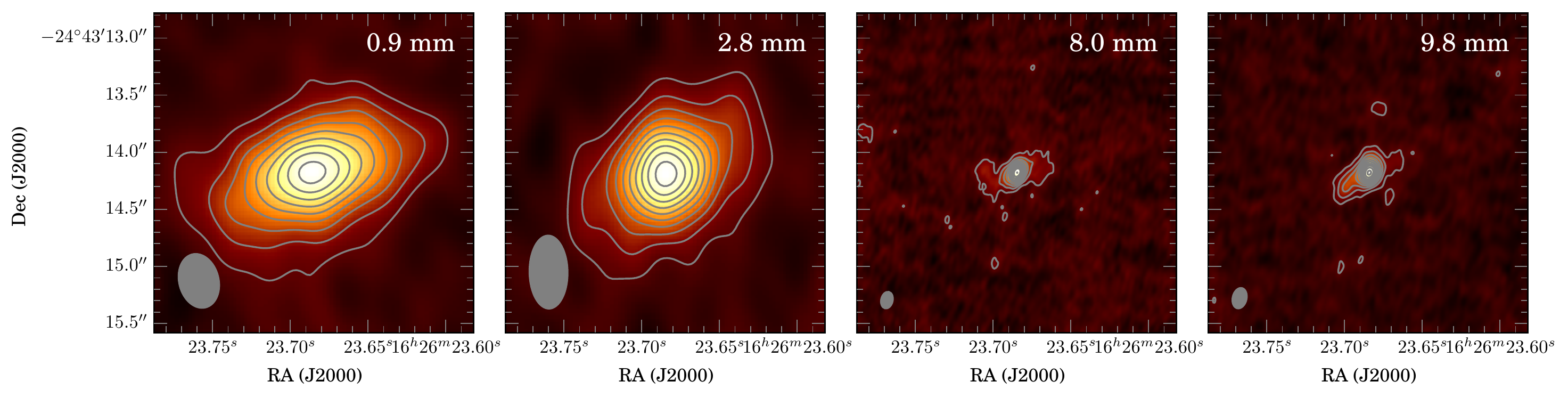}
\caption{Aperture synthesis images of the continuum emission towards the young star \doardisk, observed at wavelengths of 0.9, 2.8, 8.0, and 9.8~mm. Each panel encompasses a $2.8''\times2.7''$ region (350~AU in size at the adopted distance), with contours drawn at  $3\sigma$ intervals, where $\sigma$ is the RMS noise level in each map (see Table \ref{table_image_doar25}).}
\label{maps_doar25}
\end{center}
\end{figure*}

Synthesized maps at 1.3, 2.8, and 7.1 mm of the continuum emission from \cytau\ are presented in Figure~\ref{maps_cytau}. Each map extends $2.5''\times2.5''$, corresponding to 350~AU at the adopted distance. 
At wavelengths of 1.3~mm, 7.1~mm, and 5.0~cm, natural weighting was used to maximize the sensitivity.  At a wavelength of 2.8~mm, a Briggs weighting scheme with a robust parameter of 0.5 was used in CASA, to optimize a combination of resolution and sensitivity.  The resulting image properties and source photometry can be found in Table~\ref{table_image_cytau}, from these measurements we infer an spectral index for \cytau\ from 1.3 to 7.1~mm of $\alpha=2.6$.

The observations at 5~cm were used to estimate the contribution from processes other than thermal dust emission at 7.1~mm \citep[e.g., chromospheric activity or thermal bremsstrahlung from photoevaporative disk winds driven by the central protostar; ][]{1993Mundy,2011Pascucci}, which needs to be taken into account when modeling the dust emission from the disk.  
However, \cytau\ is not detected at 5~cm, thus we take the $3\sigma$ upper limit of 20~$\mu$Jy and estimate the maximum possible contamination at 7.1~mm by assuming optically-thick free-free emission with spherical symmetry, for which the emission as a function of frequency, $\nu$, is proportional to $S_{\nu}\propto\nu^{0.6}$ \citep[][]{1986Reynolds}.
The maximum contamination at 7.1~mm thus corresponds to $\sim 60$~$\mu$Jy, which represents only 4\% of the total emission at this wavelength. 
We note that the probable source of any such contamination will arise from very near the protostar, and would appear as an unresolved point source in the 7.1~mm data. Based on the 7.1~mm data themselves, using {\it uv}-distances $\geq 10$~km (corresponding to spatial scales smaller than $\sim0.14''$), we constrain any unresolved point source to have a flux density of $74 \pm 9$~$\mu$Jy (some of which could be dust emission), consistent with our estimate of the potential contamination extrapolated from 5~cm.

\subsection{DoAr 25}
\label{doar_obs_res}

Synthesized maps at 0.9, 2.8, 8.0 and 9.8 mm of the continuum emission from \cytau\ are presented in Figure~\ref{maps_doar25}. 
Each map extends $2.8''\times2.8''$, corresponding to 350~AU at the adopted distance. 
Natural weighting was used in the imaging at wavelengths of 8.0~mm, 9.8~mm, and 5.0~cm, to maximize the sensitivity.  
Briggs weighting with robust parameters of 0.7 and 0.3 were used for imaging at wavelengths of 0.9 and 2.8~mm, respectively, to optimize the resolution and sensitivity at these wavelengths.
The resulting image properties and source photometry can be found in Table~\ref{table_image_doar25}, from these measurements we infer an spectral index for \doardisk\ from 0.9 to 9.8~mm of $\alpha=2.8$.

\begin{deluxetable*}{c c c c c c c}
\tabletypesize{\footnotesize}
\tablecaption{\cytau\ imaging and photometry results}
\tablewidth{0pc}
\tablecolumns{7}
\tablehead{
\colhead{ }	& \colhead{ } & \multicolumn{3}{c}{Image properties} & \phantom{a} & \multicolumn{1}{c}{Source photometry\tablenotemark{a}} \\
\cmidrule{3-5}
\cmidrule{7-7}
\colhead{Telescope}	& \colhead{$\lambda$}	& \colhead{Image rms noise}	& \colhead{Synthesized beam}& \colhead{Beam P.A.}	& \phantom{a}& \colhead{Flux density} \\
\colhead{ }	& \colhead{[mm]} &\colhead{[$\mu$Jy/beam]}	& \colhead{ } & \colhead{[$^{\circ}$]} & \colhead{} & \colhead{ [mJy]} 
}
\startdata
CARMA	& 1.3 & 580	& $0.29''\times 0.24''$ & \phantom{$-$}83.3 && $119  \pm 18$   \\
CARMA	& 2.8 & 280	& $0.40''\times 0.35''$ &           $-87.1$ && $26.1 \pm 3.9$   \\
VLA		& 7.1 & 7.8 & $0.07''\times 0.07''$ &           $-49.5$ && \phantom{.}$1.76 \pm 0.27$ \\
VLA		& 50  & 7.0 & $0.62''\times 0.34''$ &           $-63.7$ && $<0.020\:(3\sigma)$
\enddata
\tablenotetext{a} {Source photometry obtained from the measured flux density of \cytau\ limited to \emph{uv}-distances between 0--40~k$\lambda$, corresponding to the shortest \emph{uv}-spacings we have sampled, except for $\lambda=50$~mm which is not detected and thus we can only provide an upper limit. Note that the errors quoted include the uncertainty in the absolute flux density scale.}
\label{table_image_cytau}
\end{deluxetable*}

\begin{deluxetable*}{c c c c c c c}
\tabletypesize{\footnotesize}
\tablecaption{\doardisk\ imaging and photometry results}
\tablewidth{0pc}
\tablecolumns{7}
\tablehead{
\colhead{ }	& \colhead{ } & \multicolumn{3}{c}{Image properties} & \phantom{a} & \multicolumn{1}{c}{Source photometry\tablenotemark{a}} \\
\cmidrule{3-5}
\cmidrule{7-7}
\colhead{Telescope}	& \colhead{$\lambda$}	& \colhead{Image rms noise}	& \colhead{Synthesized beam}& \colhead{Beam P.A.}	& \phantom{a}& \colhead{Flux density} \\
\colhead{ }	& \colhead{[mm]} &\colhead{[$\mu$Jy/beam]}	& \colhead{ } & \colhead{[$^{\circ}$]} & \colhead{} & \colhead{ [mJy]} 
}
\startdata
SMA	    & 0.9 & 3500 & $0.48''\times 0.35''$ & \phantom{$-$}14.4 && $515  \pm 52$	 \\
CARMA	& 2.8 & 270  & $0.64''\times 0.33''$ &  \phantom{$-$}0.9 && $28.1 \pm 4.2$   \\
VLA		& 8.0 & 11.0 & $0.15''\times 0.10''$ &           $-13.2$ && $1.17 \pm 0.13$  \\
VLA		& 9.8 & 8.0  & $0.18''\times 0.12''$ &           $-13.3$ && $0.66 \pm 0.07$  \\
VLA		& 50  & 5.9  & $0.68''\times 0.34''$ &            $-8.4$ && $0.051\pm 0.012$ 
\enddata
\tablenotetext{a} {Source photometry obtained from the measured flux density of \cytau\ limited to \emph{uv}-distances between 0--40~k$\lambda$, corresponding to the shortest \emph{uv}-spacings we have sampled except for $\lambda=50$~mm, where the source is weak and unresolved and the fit is performed in the image domain. Note that the errors quoted include the uncertainty in the absolute flux density scale.}
\label{table_image_doar25}
\end{deluxetable*}

As for \cytau, the observations at 5~cm from \doardisk\ were used to estimate the contribution from other than thermal dust emission at 8.0 and 9.8~mm. In the case of \doardisk, $\lambda=5$~cm emission is detected at a level of 6$\sigma$ coincident with the star, with an integrated flux density of $50\pm13$~$\mu$Jy.  Assuming $S_{\nu}\propto\nu^{0.6}$ results in an estimated contamination of $150\pm40$~$\mu$Jy at 8.0~mm, and $133\pm35$~$\mu$Jy at 9.8~mm, which corresponds to 17\% and 24\% of the integrated emission at these wavelengths.
A lower level of contamination is derived based only in the 8.0 and 9.8~mm data with {\it uv}-distances $\geq 10$~km (corresponding to spatial scales smaller than $\sim0.17''$): an unresolved point source flux density of $68\pm24$~$\mu$Jy is present in the 8.0~mm observations, while a flux density of $65\pm16$~$\mu$Jy is constrained at 9.8~mm. These estimates are consistent (within the errorbars) with the extrapolation from 5~cm. However, we adopt the most conservative value (the highest possible contamination) in our analysis.

\subsection{Brightness temperature and optical depth of the emission}

The brightness temperature of the emission at mm and cm wavelengths can be useful to discriminate between optically thin and optically thick emission.
For a medium at a physical temperature $T$ with no background radiation and an optical depth $\tau$, the brightness temperature ($T_B$) of the emission at a particular wavelength $\lambda$ is related to $T$ as: $T_B(\lambda) = T(1-e^{-\tau_{\lambda}})$ \citep[see, e.g., ][]{2009ToolsOfRadioAstronomy}.
Thus, in the optically thick limit ($\tau \gg 1$) the brightness temperature directly traces the temperature of the medium and is independent of wavelength: $T_B(\lambda)  = T$, while in the optically thin limit ($\tau \ll 1$) the brightness temperature will be much lower than the physical temperature of the medium: $T_B(\lambda) = \tau_{\lambda} T$ and will vary with wavelength according to the wavelength dependence of $\tau_\lambda$. 

\begin{figure}
\begin{center}
\hspace{0.5cm}{\bf \textsf{CY Tau}}
\includegraphics[scale=0.42]{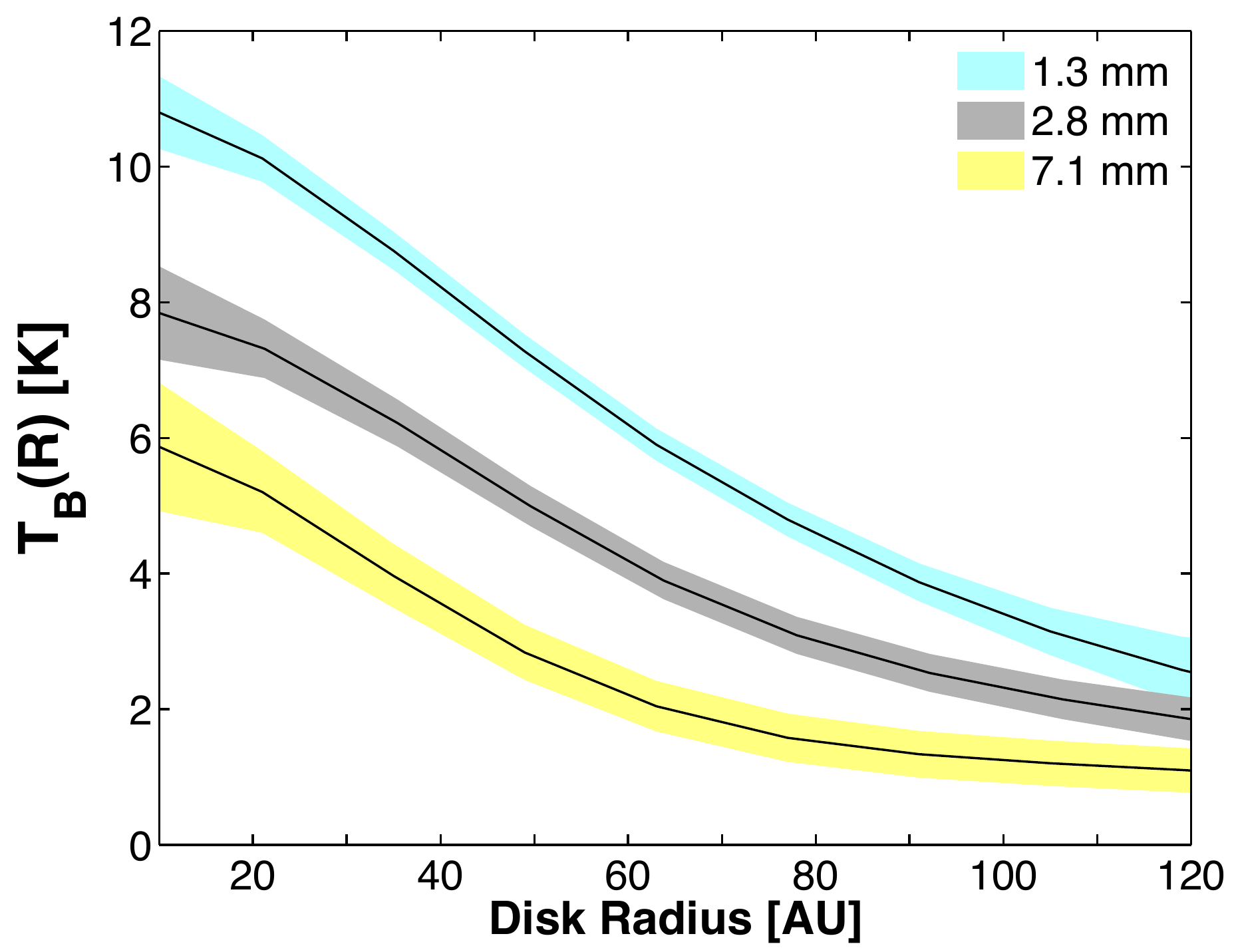}
\caption{Observed brightness temperature of the emission from our multi-wavelength observations of \cytau, as a function of the radial distance to the central star. These profiles are being compared at the same angular resolution, corresponding to the 2.8~mm observations ($0.4''$, 56~AU).}
\label{Tb_profile_cytau}
\end{center}
\end{figure}

\begin{figure}
\begin{center}
\hspace{0.5cm}{\bf \textsf{DoAr 25}}
\includegraphics[scale=0.42]{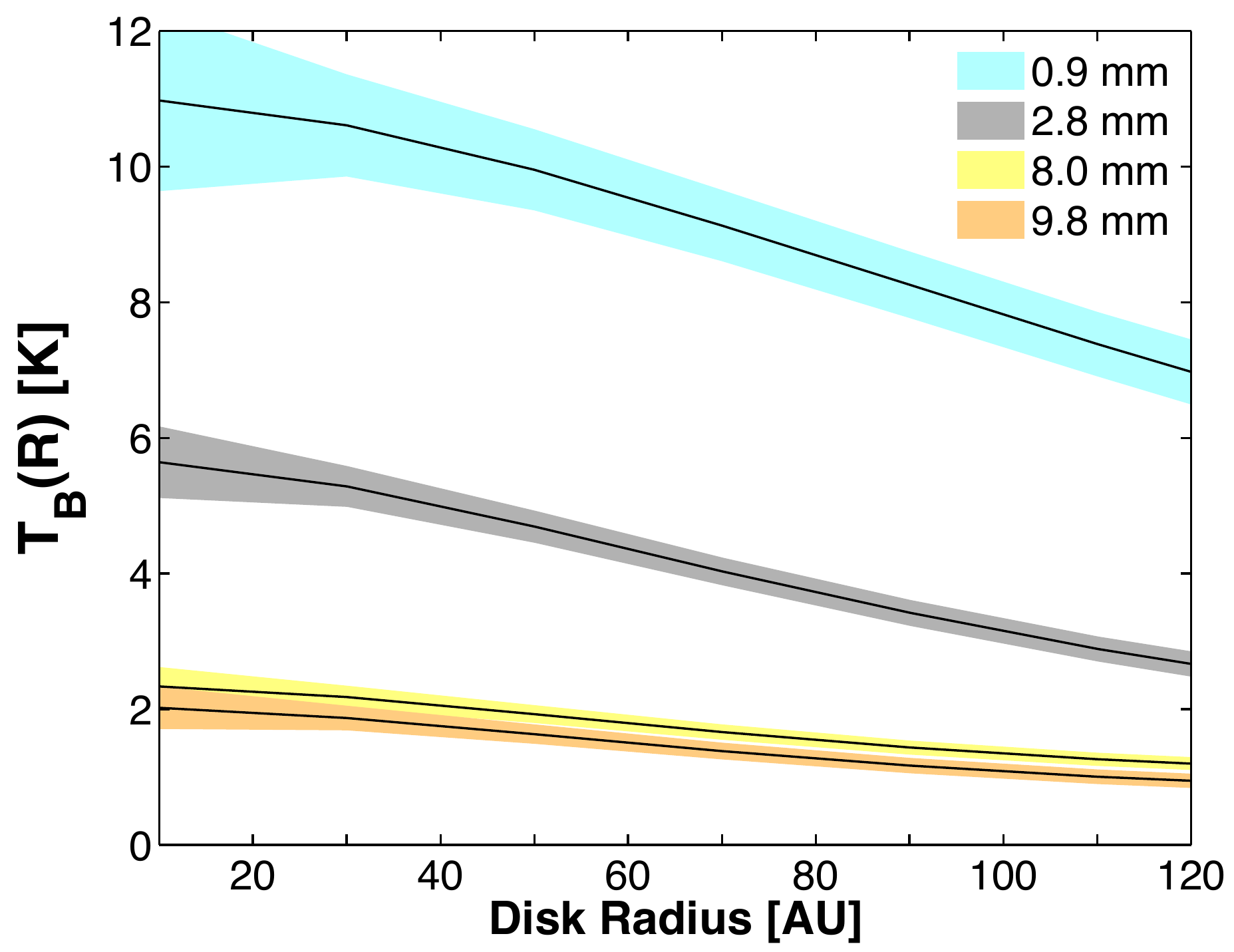}
\caption{Observed brightness temperature of the emission from our multi-wavelength observations of \doardisk, as a function of the radial distance to the central star. These profiles are being compared at the same angular resolution, corresponding to the 2.8~mm observations ($0.64''$, 80~AU).}
\label{Tb_profile_doar25}
\end{center}
\end{figure}

The data presented in \S3.1 and \S3.2 include high frequency observations and emission from the cold outer disk of \cytau\ and \doardisk, thus the Rayleigh-Jeans limit of $h\nu \ll kT$ might not be appropriate even at cm wavelengths. For this reason we calculate the brightness temperature of the emission using the full Planck function without this approximation.
Figures \ref{Tb_profile_cytau} and \ref{Tb_profile_doar25} present azimuthally-averaged radial $T_B$ profiles for both disks at each observed wavelength, with the shaded region representing the $1\sigma$ constraint derived from the statistical uncertainty in the images. For a proper comparison, we convolved all observations to the same angular resolution (i.e., the lowest spatial resolution, which corresponds to the resolution of the 2.8 mm data).  In addition, we subtracted an unresolved point-source component for the long-wavelength observations at 7.1~mm (\cytau) and at 8.0 and 9.8~mm (\doardisk), whose flux density is given by the amount of contamination from non-dust emission described in \S3.1 and \S3.2. 

The observed brightness temperature profiles in Figures \ref{Tb_profile_cytau} and \ref{Tb_profile_doar25} decline with increasing distance from the star in both disks, becoming successively fainter as the wavelength of the emission is increased. 
If the observed emission were optically thick, the $T_B$ profiles would directly trace the physical temperature of the disk. For a vertically isothermal disk in this limit, the observed brightness temperature should then be the same at all wavelengths. 
For both \doardisk\ and \cytau\ we observe a successively fainter $T_B$ profile with increasing wavelength, suggesting that for all but the shortest wavelength, the observed emission is consistent with being optically thin. 
Furthermore, in \S5 we find that, for a flared disk in hydrostatic equilibrium, the midplane temperature is a factor of several higher than the observed brightness temperature of \cytau\ and \doardisk\ at {\it all} disk radii and at {\it all} wavelengths, thus indicating that the observed emission is optically thin rather than optically thick (see section \S4 for details of the modeling and \S5.1 for the resulting midplane temperature profiles).

A colder midplane temperature could be found in the extreme case of a thin flat disk in the presence of no interstellar radiation. 
This configuration results in the coldest midplane disk possible, since the opening angle of a thin disk is much smaller than for a flared disk, making the stellar heating less efficient. 
Note that a more settled disk interpretation is favored in the analysis of unresolved observations of \cytau, as the SED from mid-infrared to far-infrared wavelengths is quite steep for this object \citep[see, e.g., ][]{2007ApJS..169..328R}.
In the case of \doardisk, Spitzer IRS spectra reveal a flat spectrum between 10--30 microns consistent with a flared disk structure \citep{2009A&A...507..327O,2010ApJS..188...75M}. 
Thus, we only consider the thin disk case for \cytau.  We find a midplane temperature of $\sim 14$~K at 10~AU, decreasing to a lower limit defined by interstellar radiation and cosmic ray heating (assumed to be $\sim 8$~K) at radii of 30 AU and greater. Only for this extreme case is the observed $T_B$ profile of \cytau\ consistent with the physical temperature of the disk midplane, and this is only for the shortest wavelength observations at 1.3~mm at radii $\sim$30 to 50 AU. At longer wavelengths (and at larger radii at 1.3~mm) the brightness temperature is still below this cold midplane case, which indicates that the optical depth of the emission is less than 1, even in this extreme case.
Consequently, we will assume in the following analysis that the emission at 1.3~mm from \cytau\ is optically thin, although there is a possibility of higher optical depth at 1.3~mm if the disk structure corresponds to a flat thin disk for this object.

\subsection{De-projected visibility profiles}

\begin{figure}
\begin{center}
\includegraphics[scale=0.42]{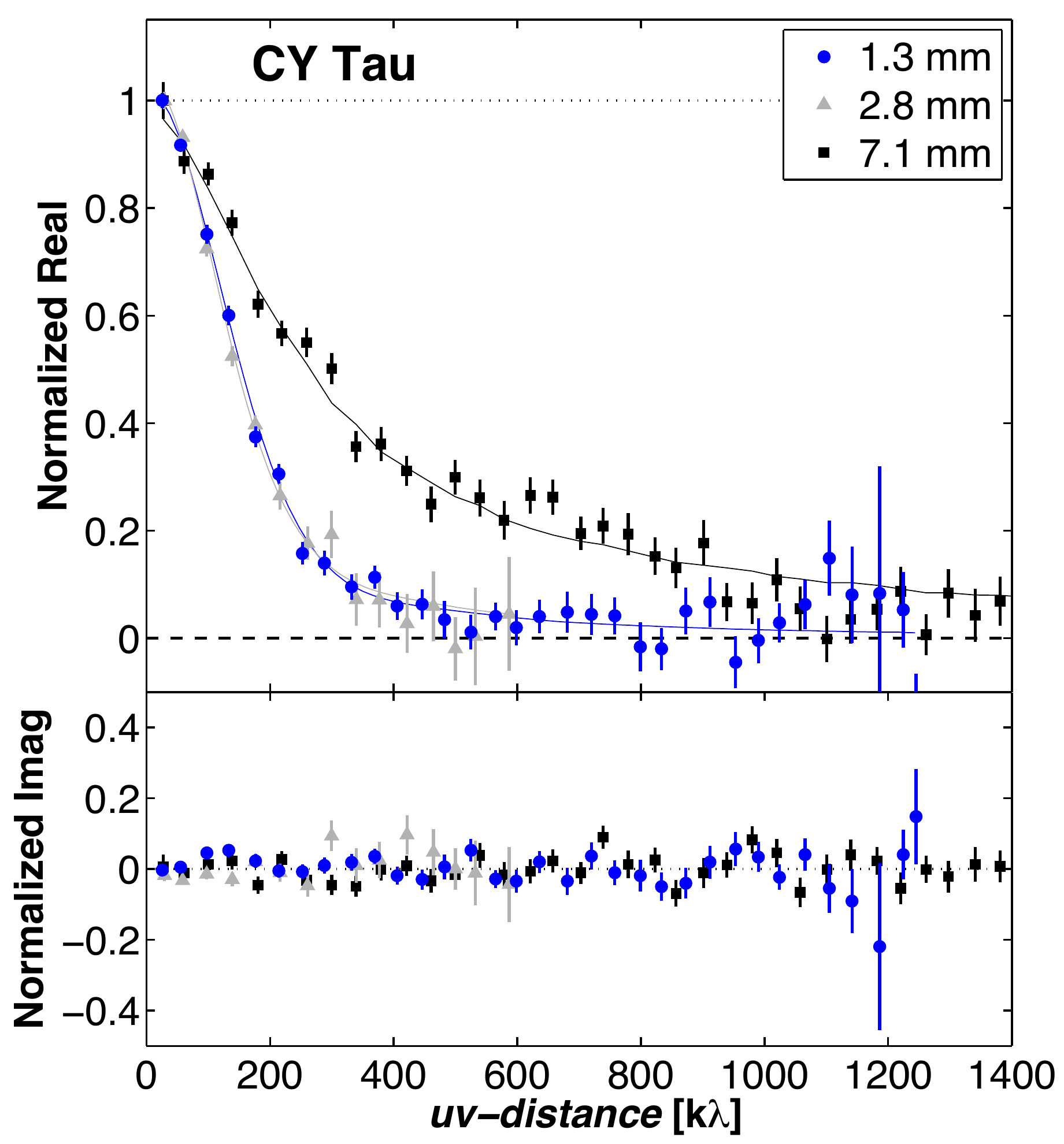}
\caption{Normalized real and imaginary part of the correlated emission from \cytau\ as a function of spatial frequency (\emph{uv}-distance). 
The visibilities have been de-projected  using a position angle of $63^{\circ}$ and inclination of $28^{\circ}$, derived from molecular line observations \citep[][]{2011Guilloteau}. Each bin has a width of 40~k$\lambda$, and has been normalized by the measured flux density at 0--40~k$\lambda$. Filled circles and error bars of different color correspond to correlated real and imaginary part of the emission observed at different wavelengths (1.3~mm: blue circle, 2.8~mm: gray triangle, 7.1~mm: black square). The continuous lines correspond to the best-fit disk emission model at each wavelength (see modeling in \S5.1).}
\label{vis_profile_cytau}
\end{center}
\end{figure}

\begin{figure}
\begin{center}
\includegraphics[scale=0.42]{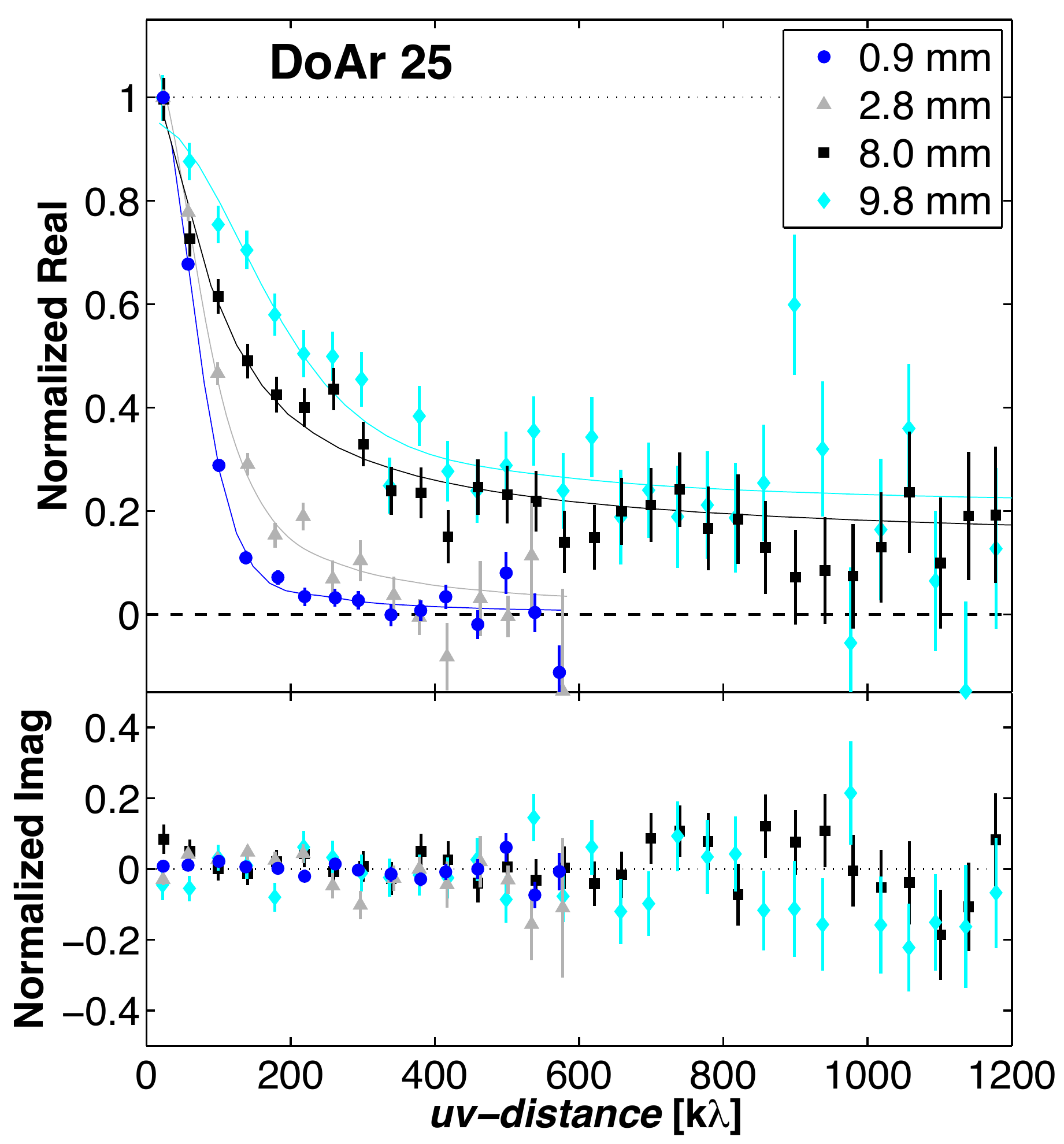}
\caption{
Normalized real and imaginary part of the correlated emission from \doardisk\ as a function of spatial frequency (\emph{uv}-distance). 
The visibilities have been de-projected assuming a position angle of $109^{\circ}$ and inclination of $62^{\circ}$, as derived from the modeling of the shortest-wavelength emission. Each bin has a width of 40~k$\lambda$, and has been normalized by the measured flux density at 0--40~k$\lambda$. Filled circles and error bars of different color correspond to the correlated real and imaginary parts of the emission observed at different wavelengths (0.9~mm: blue circle, 2.8~mm: gray triangle, 8.0~mm: black square, 9.8~mm: cyan diamond). The continuous lines correspond to the best-fit disk emission model at each wavelength (see modeling in \S5.1).}
\label{vis_profile_doar25}
\end{center}
\end{figure}

In Figures \ref{vis_profile_cytau} and \ref{vis_profile_doar25}, we present the real and imaginary part of the visibility as a function of \emph{uv}-distance (hereafter referred to as visibility profiles), for both \cytau\ and \doardisk, at each of the observed wavelengths.
Each visibility was deprojected by the known inclination and position angle of the disk (see Section~\ref{modeling_emission} for details of the known disk geometry), before averaging into {\it uv}-bins with a width of 40k$\lambda$. 
To compare the different wavelengths, each visibility profile is normalized by the measured flux density in the first \emph{uv}-bin, between 0--40~k$\lambda$.
In Figures \ref{vis_profile_cytau} and \ref{vis_profile_doar25}, the black dotted line at a constant value of $Re=1.0$ represents the observed visibility profile for an unresolved point source. Once a disk is resolved by interferometric observations, its visibility profile will decline from this reference line. 
Thus, the significant decline of the real part of the visibility profile at each wavelength demonstrates that the emission from both \doardisk\ and \cytau\ is resolved, at all the observed wavelengths. 
Furthermore, a compact source will have a shallower decline with \emph{uv}-distance than an extended source, whose decline will be quite steep. 
Thus, since the decline of the short-wavelength visibility profile (0.9, 1.3, and/or 2.8 mm) is steeper than the long-wavelength visibility profile (7.1, 8.0 and/or 9.8~mm),  the disk emission observed at short-wavelengths is more extended than the disk emission observed at long-wavelengths, which has to be more compact. Hence, the differing visibility profiles demonstrate a wavelength-dependent structure in both the \cytau\ and \doardisk\ disks, similar to that found for the disks surrounding AS~209 \citep{2012Perez}, CQ~Tau \citep{2011Banzatti}, and several other stars in the Taurus-Auriga star-forming region \citep{2011Guilloteau}. In the next section, we explain the wavelength-dependent structure of the emission from \cytau\ and \doardisk\ as radial variations of the dust properties across the circumstellar disk.

\section{Modeling of the disk emission}
\label{modeling_emission}

Observations of \cytau\ and \doardisk\ were analyzed using a disk emission model that reproduces the radial brightness distribution at mm and cm wavelengths, equivalent to the disk model employed in the analysis of similar observations of the disk surrounding the young star AS~209 \citep{2012Perez}. 
At $\lambda= 7.1$, 8.0, and 9.8~mm an additional point source at the center of the disk is also included, to represent the potential contamination from sources other than thermal dust emission, as described in Sections~\ref{cytau_obs_res} and \ref{doar_obs_res}. 
We adopt the two-layer disk approximation for the disk structure, first presented by \citet{1997Chiang}. 
In this model, a flared, passively-heated disk is irradiated by the central star, with its structure characterized by two components: a disk surface layer that absorbs the stellar radiation, and a disk interior, which is opaque to the stellar photons. 
Dust in the surface layer will absorb the short-wavelength stellar emission and re-radiate it at longer wavelengths.
As the disk interior absorbs the reprocessed emission, the inner regions of the disk can be efficiently heated.
We use the implementation of this radiative transfer problem as described by \citet{2009Isella}, where the temperature structure of the two-layer model is computed with the iterative method presented by \citet{2001Dullemond}. 
The assumptions implicit in the two-layer disk model break down for the far outer regions of the disk, where the surface density of material drops significantly and even the disk interior becomes optically thin to the stellar radiation. 
When this condition is met, we assume the same temperature for both the surface layer and disk interior.
The disk temperature smoothly decreases as a power-law function of radius down to a minimum of 10~K in the outer disk, where the dust is heated by the interstellar radiation field.

For the mass surface density structure, $\Sigma(R)$, we adopt the similarity solution for a viscously-accreting Keplerian disk subject to the gravity of a massive central object \citep{1974LBP}, which can be described by a power law combined with an exponential taper at large radii. If the disk viscosity, $\nu_v$, is assumed to be a power-law with radius ($\nu_v \propto R^{\gamma}$), then the solution for the disk surface density is time-independent \citep{1998ApJ...495..385H}. We adopt the same parameterization of the similarity solution presented by \citet{2011Guilloteau}:
\begin{equation}
\Sigma(R) = \Sigma_0 \left( \frac{R}{R_0} \right)^{-\gamma} \times \exp \left( - \left(\frac{R}{R_C}\right)^{2-\gamma} \right)
\label{Similarity}
\end{equation}
where $\Sigma_0$ corresponds to the surface density at radius $R_0$, $\gamma$ is the power-law exponent on the radial dependence of the viscosity, and $R_C$ is the characteristic scaling radius\footnote{The scaling radius $R_C$ is directly related to the location in the disk where the mass accretion rate changes sign, the transition radius: $R_t = R_C \left(\frac{1}{2(2-\gamma)}\right)^{1/(2-\gamma)}$.
The transition radius defines a boundary for accretion or expansion, since for $R < R_t$ mass flows inward while for $R > R_t$ mass flows outward. Note that in the context of viscous disk evolution, the value of the transition radius will increase with time due to conservation of angular momentum: as matter is accreted onto the star the disk must expand to conserve total angular momentum. We include its definition here since different authors will either use $R_C$ or $R_t$ in their preferred prescription for the disk surface density.}. This prescription for $\Sigma(R)$ behaves as a power law for small radii, while at large radii it decreases smoothly in an exponential fashion.

The inner radius of the disk, $R_{in}$, is defined as the radius at which the dust sublimates: 
 \begin{equation}
R_{in} =  0.07~{\rm AU} \left( \frac{L_{\star}}{L_{\odot}} \right)^{1/2} \left( \frac{T_{subl}}{1500~{\rm K}} \right)^{-2}
\end{equation}
which corresponds to $\sim0.04$~AU for \cytau\ and $\sim0.06$~AU for \doardisk, for a dust sublimation temperature of  $T_{subl}=1500$~K. 
Given that the angular resolution of our observations is more than an order of magnitude greater than $R_{in}$, the value of this parameter does not affect our constraints on the disk structure.

As in the analysis presented by \citet{2012Perez}, we compute the magnitude of the dust opacity, $\kappa_{\lambda}$, using Mie theory. 
We adopt a grain population of compact spherical grains larger than $a_{min}=0.01~\mu$m, in a power-law distribution of grain sizes: $n(a) \propto a^{-q}$ for $a_{min}<a<a_{max}$. 
Grains in the disk interior are assumed to be composed of silicates, organic materials, and water ice, while grains in the disk surface are assumed to be depleted of ice. 
We obtain optical constants for these grain materials from \citet{2003Semenov} (for silicates), \citet{1996Zubko} (for amorphous carbon), and \citet{1984Warren} (for water ice). 
Recent studies have revised the solar abundance of oxygen \citep{2009Asplund}, so although we assume the fractional abundances of these materials as given by \citet{1994Pollack}, we reallocate the fractions of these astrophysical grains to account for the increased oxygen abundance. 
This results in grains in the disk interior which are composed of 12\% silicates, 44\% organics and 44\% water ice, while in the disk surface (where water ice may not be present) the fractional abundances correspond to 21\% silicates and 79\% organics. 
Except for the normalization of the surface density at radius $R_0$ ($\Sigma_0$), the effect of dust grain composition in the disk modeling will produce a minimal effect in the derived parameters of the disk structure \citep[well within their uncertainties, as shown by ][]{2010Isella}. 
More specifically, although the dust composition does affect the absolute value of opacity, the fact that the dust emission is optically thin at long wavelengths (\S5.3) means that the derivation of the dust opacity spectral index $\beta$ is insensitive to the assumed absolute opacity value and hence dust composition.
This follows since the only frequency dependence when the emission is optically thin corresponds to $S_{\nu} \propto \nu^{\beta} * B_{\nu}(T)$.

Finally, the last parameters that determine the observed disk emission describe the geometry of the disk in the plane of the sky: inclination ($i$) and position angle (PA, measured from North to East). 
In the case of \cytau, the disk geometry has been characterized from molecular line observations of CO at $0.5''$ resolution: $i=28^{\circ}\pm5^{\circ}$, PA$=63^{\circ}\pm1^{\circ}$ \citep[][]{2011Guilloteau}, which is what we adopted for this study \citep[note that these constraints are consistent with those derived from CN observations at lower spatial resolution: $i=24.0^{\circ}\pm2.4^{\circ}$, PA$=62.5^{\circ}\pm1.8^{\circ}$;][]{2014A&A...567A.117G}.
For \doardisk, there are no molecular line observations in the literature. 
We therefore constrained its disk geometry from our modeling of the dust continuum observations at 0.9~mm, where the signal-to-noise ratio is the highest, finding $i=62^{\circ}\pm1^{\circ}$ and PA$ = 109^{\circ}\pm2^{\circ}$ \citep[consistent with previous modeling by][]{2008Andrews,2009Andrews}.

We find the best-fit model to a single wavelength observation through $\chi^2$ minimization using a  Markov-Chain Monte Carlo (MCMC) procedure \citep[see ][]{2009Isella}. 
The $\chi^2$ probability distribution is sampled by varying the free parameters that define the surface density profile ($\Sigma_0, R_C, \gamma$), since the inclination and position angle of the disk are fixed to the values reported above, and where $R_0$ has been fixed to a value of $20$~AU.
A set of these parameters defines a model of the disk brightness distribution, with the addition of a point source at the disk center whose flux density corresponds to the estimated level of contamination from non-dust thermal emission, as inferred in Section \ref{ObservationalResults}. From this model emission, we produce an image of the disk making sure that the significant spatial scales are covered.
We take the Fourier transform of this image and obtain model visibilities sampled at the same locations in the \emph{uv}-domain as the original observations. 
We compare model and observed visibilities by means of the $\chi^2$, and construct 16 separate MCMC chains that sample the parameter space using a  Metropolis-Hastings algorithm with a Gibbs sampler. 

These chains all converged to the best-fit model, which  corresponds to the one that minimizes the $\chi^2$ and hence is the model that best reproduces our observations.
We note that this MCMC algorithm samples the posterior probability distribution function (PDF) of the parameters when chains have converged and reached equilibrium.  Thus, to find best-fit values and confidence intervals for the parameters in our model we adopt a Bayesian approach, marginalizing the resulting PDF over all but one parameter to obtain the probability distribution of the parameter of interest. From each parameter PDF we find the region that contains  68.3\%($1\sigma$), 95.5\%($2\sigma$), and 99.7\%($3\sigma$) of all samples at equal probability to constrain each of the aforementioned parameters.

\begin{deluxetable}{c c c c c}
\tabletypesize{\footnotesize}
\tablecaption{Best-fit model parameters and 1$\sigma$ constraints for \cytau}
\tablewidth{0pc}
\tablecolumns{5}
\tablehead{
\colhead{$\lambda$}	& \colhead{$R_C$}		& \colhead{$\gamma$}	& \colhead{$\Sigma_0$}		& \colhead{$\chi^2_{\rm red}$}\\
\colhead{$[$mm$]$}	& \colhead{[AU]}		& \colhead{}			& \colhead{[gm cm$^{-2}$]}} & \colhead{}
\startdata
1.3 		& $62.5^{+1.8}_{-1.6}$	    & $0.06^{+0.06}_{-0.06}$    & $2.36^{+0.07}_{-0.07}$    &1.15\\
2.8 		& $64.4^{+2.3}_{-2.1}$	    & $0.16^{+0.11}_{-0.07}$	& $3.32^{+0.11}_{-0.12}$    &1.14\\
7.1			& $36.0^{+2.2}_{-1.6}$      & $0.55^{+0.06}_{-0.08}$	& $6.12^{+0.36}_{-0.33}$    &1.14
\enddata
\label{table_modeling_cytau}
\end{deluxetable}

\begin{deluxetable}{c c c c c}
\tabletypesize{\footnotesize}
\tablecaption{Best-fit model parameters and 1$\sigma$ constraints for \doardisk}
\tablewidth{0pc}
\tablecolumns{5}
\tablehead{
\colhead{$\lambda$}	& \colhead{$R_C$}		& \colhead{$\gamma$}	& \colhead{$\Sigma_0$}		& \colhead{$\chi^2_{\rm red}$}\\
\colhead{$[$mm$]$}	& \colhead{[AU]}		& \colhead{}			& \colhead{[gm cm$^{-2}$]}  & \colhead{}
}
\startdata
0.9 		& $123.2^{+3.5}_{-2.8}$	    & $-0.12^{+0.05}_{-0.05}$   & $0.64^{+0.03}_{-0.03}$    &1.08\\
2.8 		& $103.5^{+5.9}_{-4.7}$	    & \phantom{$-$}$0.34^{+0.07}_{-0.08}$	& $1.21^{+0.05}_{-0.06}$    &1.07\\
8.0			& $106.0^{+38.2}_{-21.7}$   & \phantom{$-$}$0.83^{+0.08}_{-0.13}$	& $1.28^{+0.11}_{-0.08}$    &1.11\\
9.8			& $45.5^{+7.1}_{-4.2}$      & \phantom{$-$}$0.32^{+0.14}_{-0.19}$	& $2.70^{+0.31}_{-0.28}$    &1.12
\enddata
\label{table_modeling_doar25}
\end{deluxetable}

\section{Modeling results}
\label{results_modeling}

To model our \cytau\ and \doardisk\ observations, we first select the maximum grain size ($a_{max}$) and grain size distribution slope ($q$) that best reproduces the unresolved SED from sub-mm to cm wavelengths. For \cytau, these correspond to $q=3.5$ and $a_{max}=2.5~$mm, which results in an opacity spectral slope $\beta = 0.79$ between 1.3 and 7.1~mm. For \doardisk, the long-wavelength SED can be reproduced with $q=3.5$ and $a_{max}=1.5~$mm, resulting in an opacity spectral slope $\beta = 0.91$ between 0.9 and 8.0~mm. Note that while these values of $a_{max}$ and $\beta$ are consistent with the SEDs, they are relatively poorly constrained by the curvature of the long wavelength emission.

In order to minimize the number of free parameters in this model, we assume a \emph{radially constant dust opacity} throughout the disk and fit each wavelength separately. Our model then constrains the disk surface density ($\Sigma(R)$) and temperature profile ($T(R)$) at each wavelength; these constraints will appear to be different at different wavelengths if the assumption of a constant dust opacity is not satisfied.
The same approach has been successfully employed to measure radial variations of the dust opacity in the disks surrounding RY~Tau, DG~Tau, and AS~209 \citep{2010Isella,2012Perez}; in the following sections we present our results for the circumstellar disks of \cytau\ and \doardisk.

\subsection{CY Tau and DoAr 25 disk structure constraints}
\label{sigma_temp_results}

\begin{figure}
\begin{center} 
\hspace{1.5cm}{\bf \textsf{CY Tau}}
\includegraphics[scale=0.42]{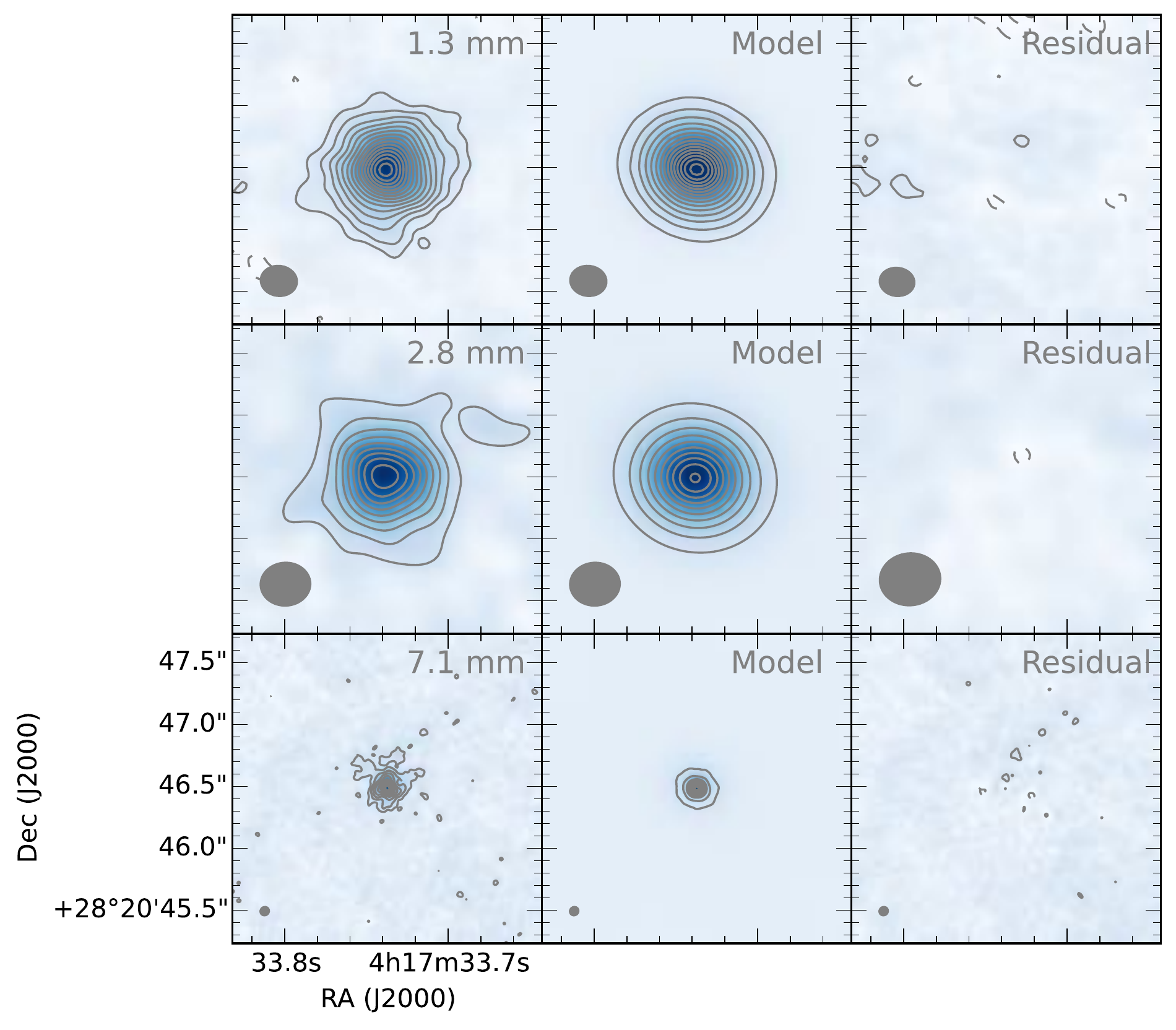}
\caption{Modeling of the 1.3~mm (top), 2.8~mm (middle), and 7.1~mm (bottom) continuum emission towards \cytau. For each row, the data are shown on the left, the best-fit model disk is shown on the middle panel, and the residual (obtained by subtracting the best-fit model from the data in the Fourier domain) is on the right. The imaging parameters are as described in Section~\ref{cytau_obs_res}. Contours start at $-3\sigma$, stepping by $3\sigma$, where $\sigma$ is the RMS noise in each map (Table \ref{table_image_cytau}).}
\label{maps_model_res_cytau}
\end{center}
\end{figure}

\begin{figure}
\begin{center} 
\hspace{1.5cm}{\bf \textsf{DoAr 25}}
\includegraphics[scale=0.42]{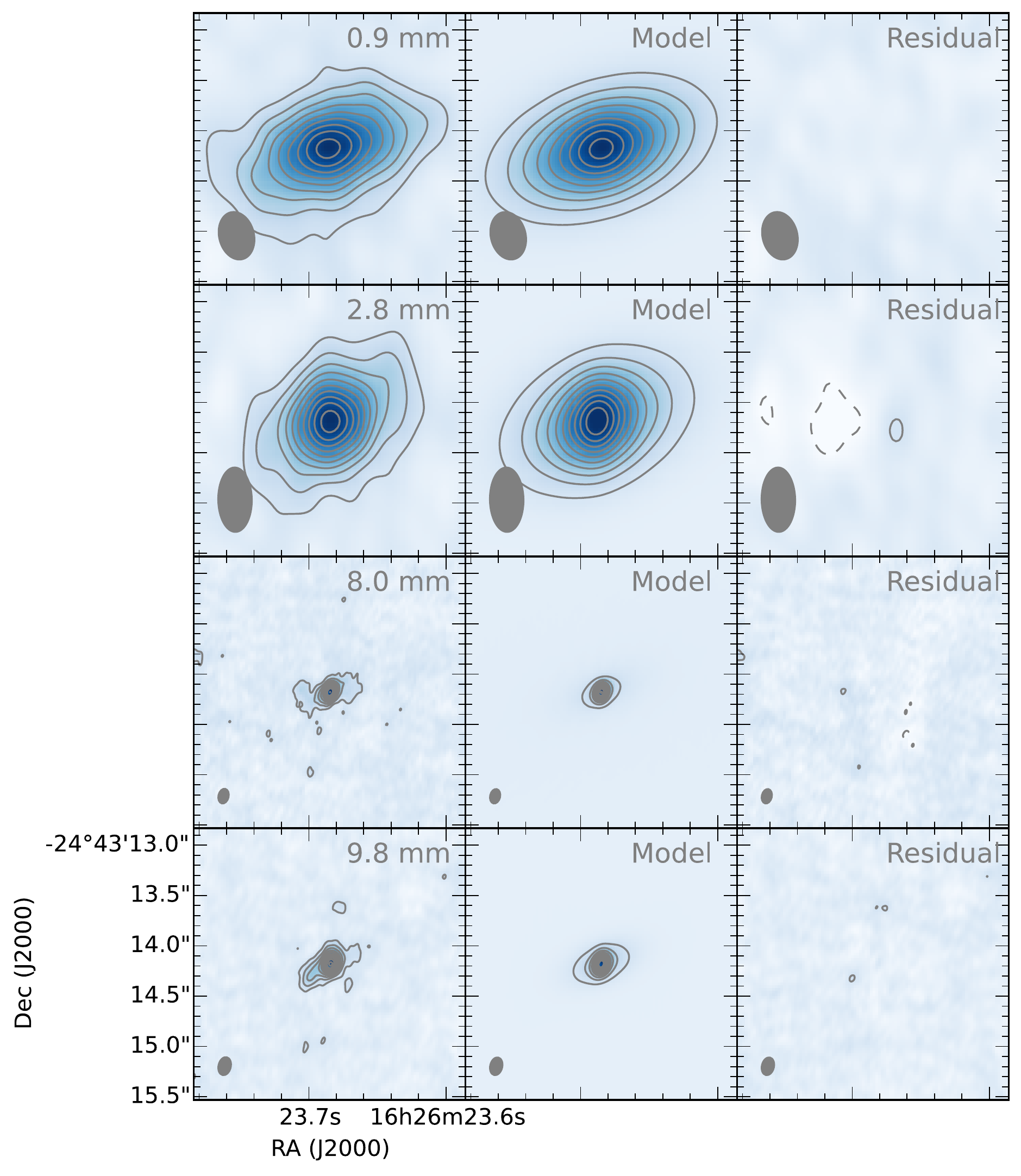}
\caption{Modeling of the 0.9~mm (top), 2.8~mm (2nd from top), 8.0 and 9.8~mm (bottom two rows) continuum emission towards \doardisk. For each row, the data are shown on the left, the best-fit model disk is shown on the middle panel, and the residual (obtained by subtracting the best-fit model from the data in the Fourier domain) is on the right. The imaging parameters are as described in Section~\ref{doar_obs_res}. Contours start at $-3\sigma$, stepping by $3\sigma$, where $\sigma$ is the RMS noise in each map (Table \ref{table_image_doar25}).}
\label{maps_model_res_doar25}
\end{center}
\end{figure}

The best-fit and $1\sigma$ constraints for each model parameter ($\Sigma_0,R_C,\gamma$), obtained from their marginalized probability distribution function, are presented in Tables \ref{table_modeling_cytau} and \ref{table_modeling_doar25}.
To demonstrate that these best-fit models are sensible representations of the observations, we imaged the model and residual visibilities (calculated by subtracting the model visibilites from the data) in the same way as the original observations.
The resulting images are shown in Figure~\ref{maps_model_res_cytau} and Figure~\ref{maps_model_res_doar25} for \cytau\ and \doardisk\ respectively. 
Additionally, for each of the best-fit models, we computed de-projected visibility profiles (solid lines in Figures~\ref{vis_profile_cytau} and \ref{vis_profile_doar25}). 
The adopted model is a good representation of the data at each separate wavelength, as no significant residual emission can be seen in the residual maps and the visibility profile for our best-fit model agrees with the observations at all wavelengths.

\begin{figure*}
\begin{center}
\includegraphics[scale=0.31]{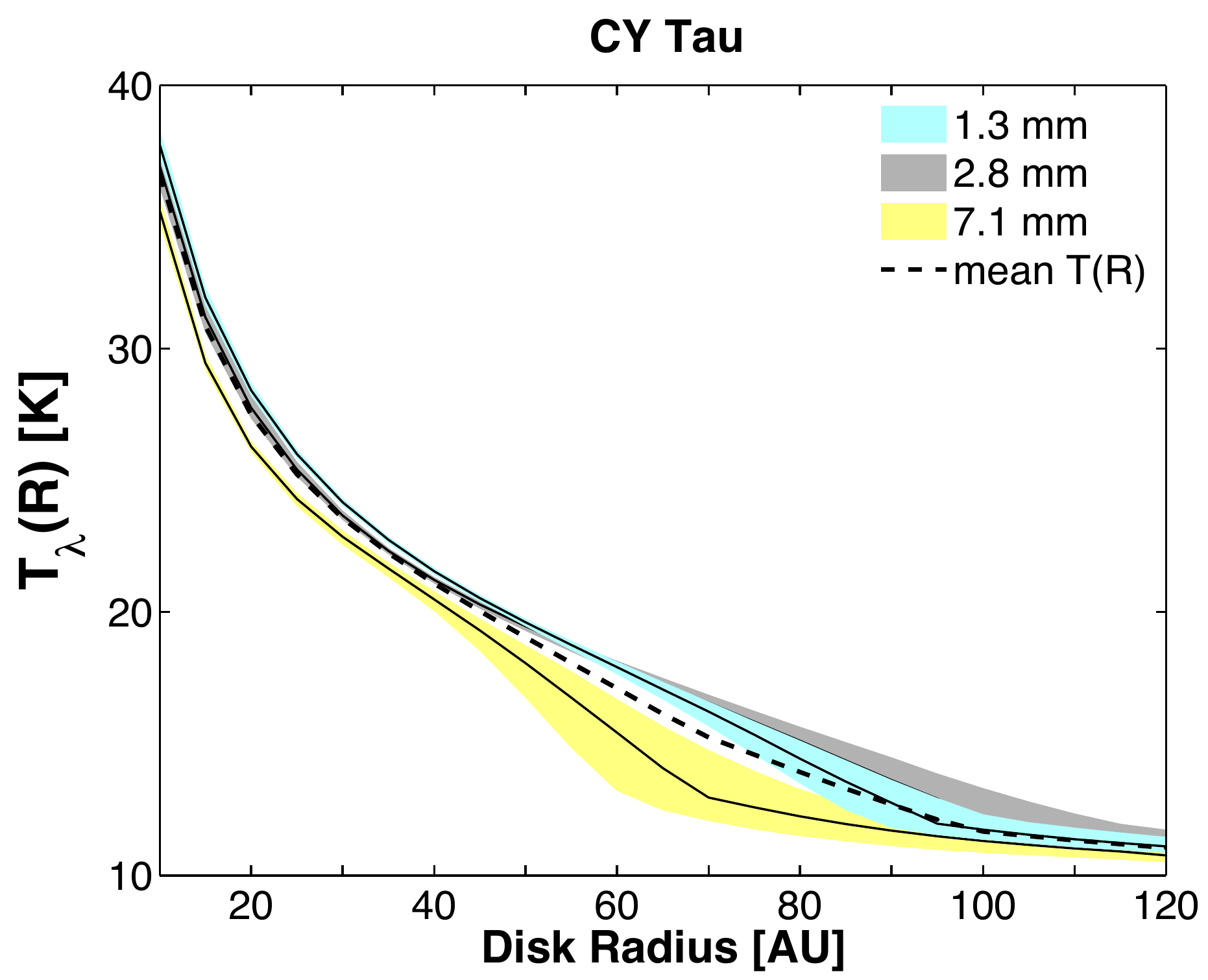}
\includegraphics[scale=0.31]{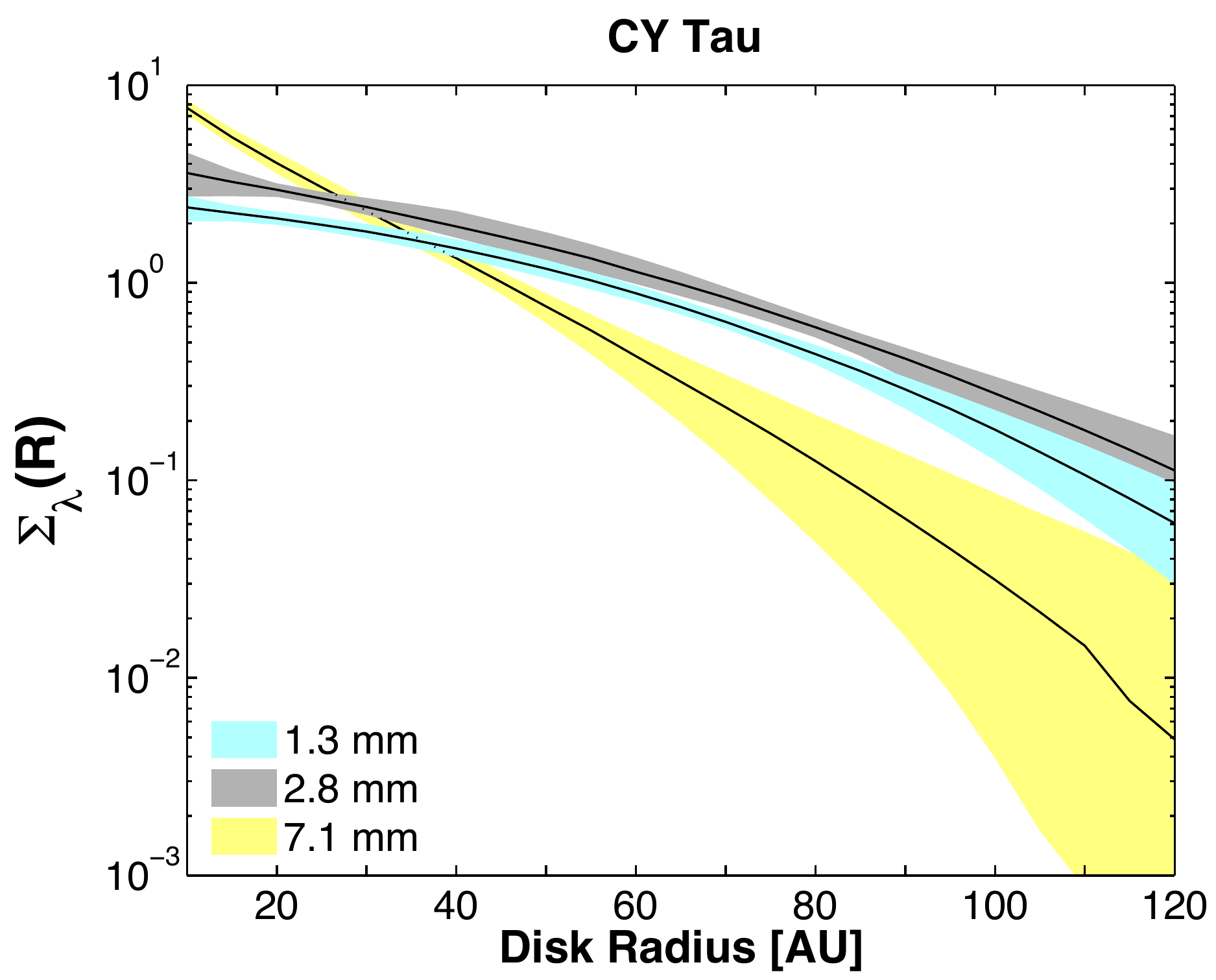}
\includegraphics[scale=0.31]{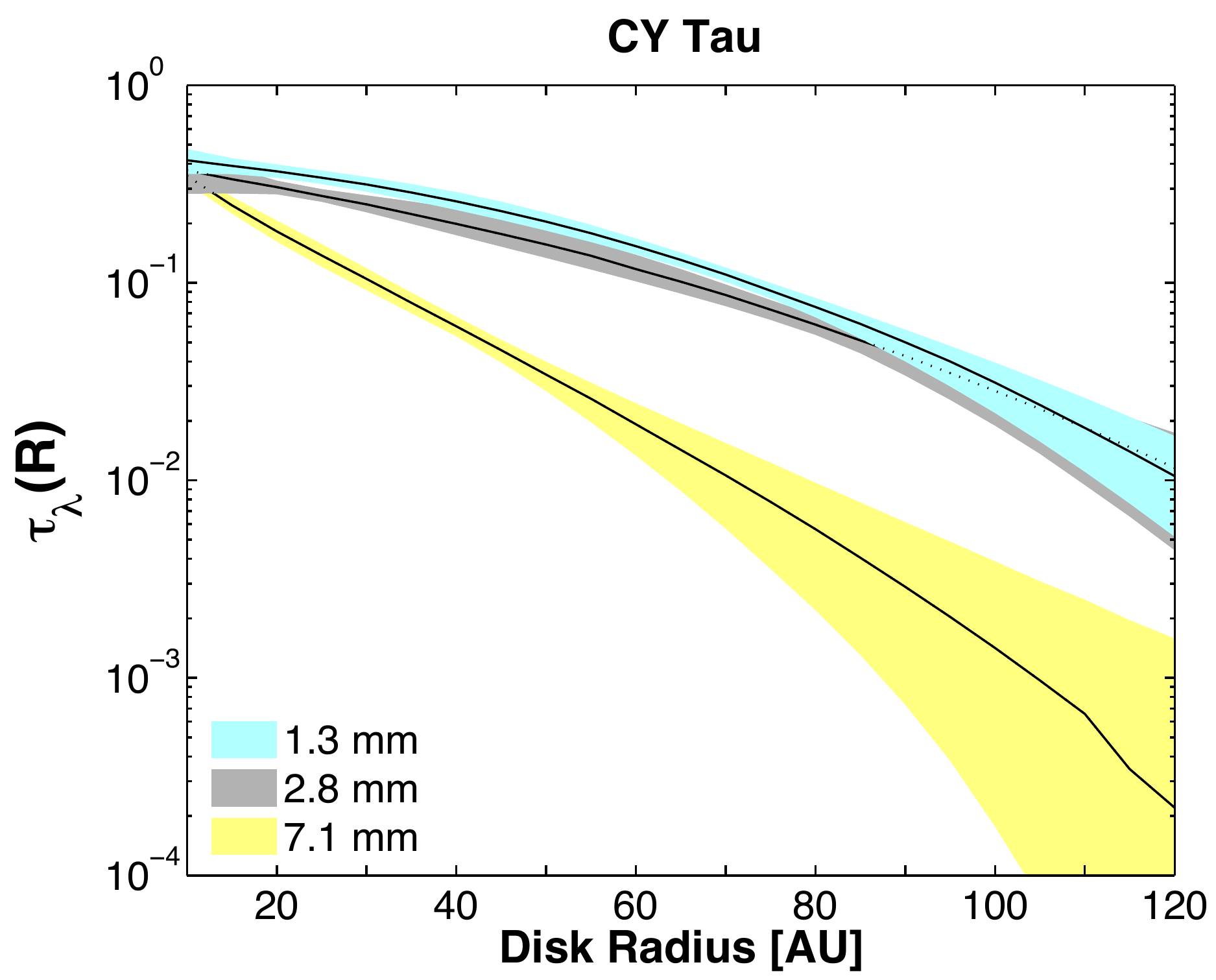}
\caption{\emph{Left and middle:} $T_{\lambda}(R)$ and $\Sigma_{\lambda}(R)$ inferred from separate modeling of our multi-wavelength observations of \cytau, assuming a constant dust opacity with radius. \emph{Right:} Optical depth $\tau_{\lambda}(R)=\kappa_{\lambda}\times\Sigma_{\lambda}(R)$ inferred from separate modeling of \cytau\ multi-wavelength observations, assuming a radially constant $\kappa_{\lambda}$.
Colored regions: $3\sigma$ confidence interval constrained by our observations; continuous line: best-fit model; dashed line on left panel: mean temperature profile. The different $\Sigma(R)$ and $T(R)$ profiles for each wavelength imply a varying dust opacity with radius, and because of this, none of them is the true surface density and temperature profile of the disk.}\label{CYTau_structure}
\end{center}
\end{figure*}

\begin{figure*}
\begin{center}
\includegraphics[scale=0.31]{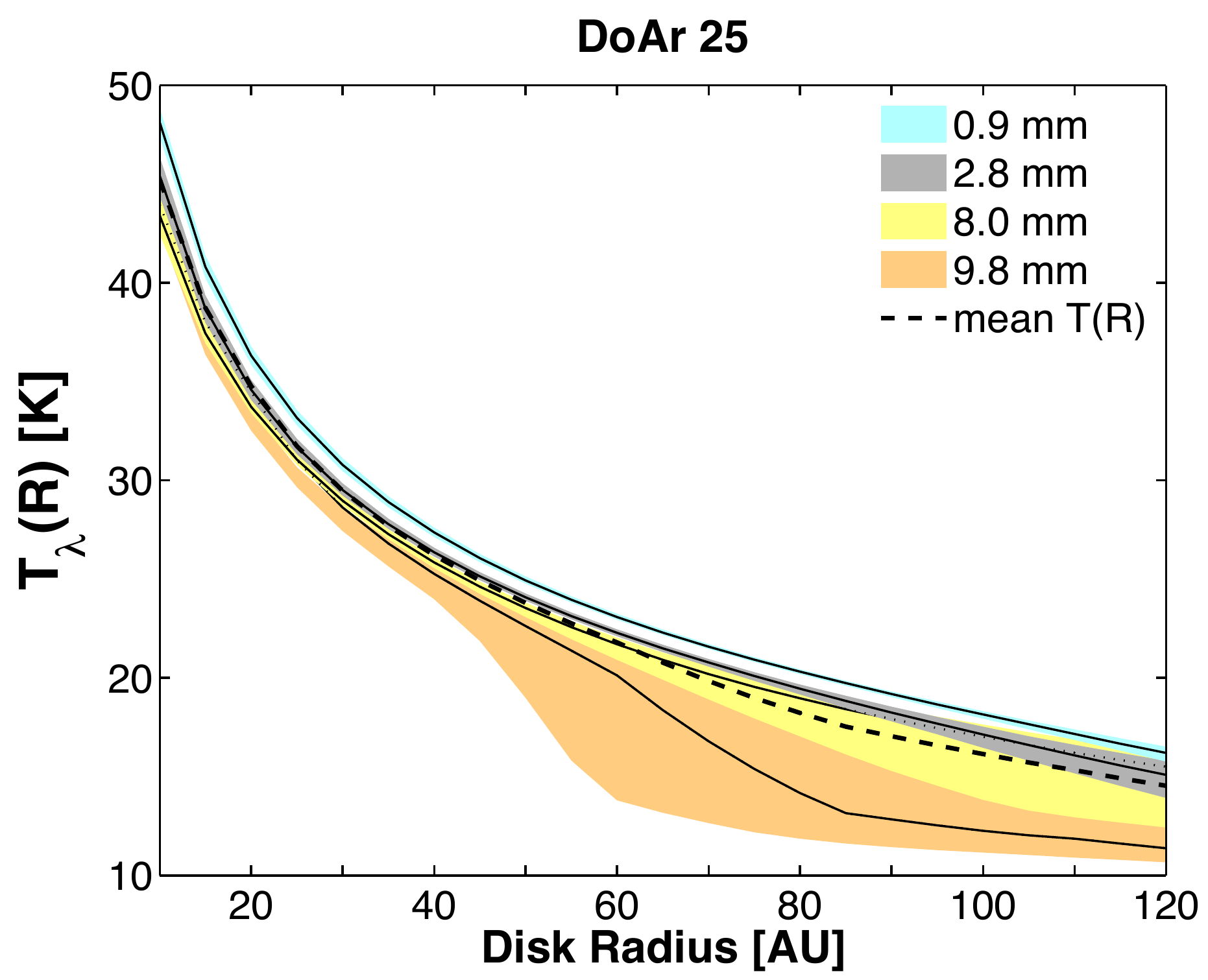}
\includegraphics[scale=0.31]{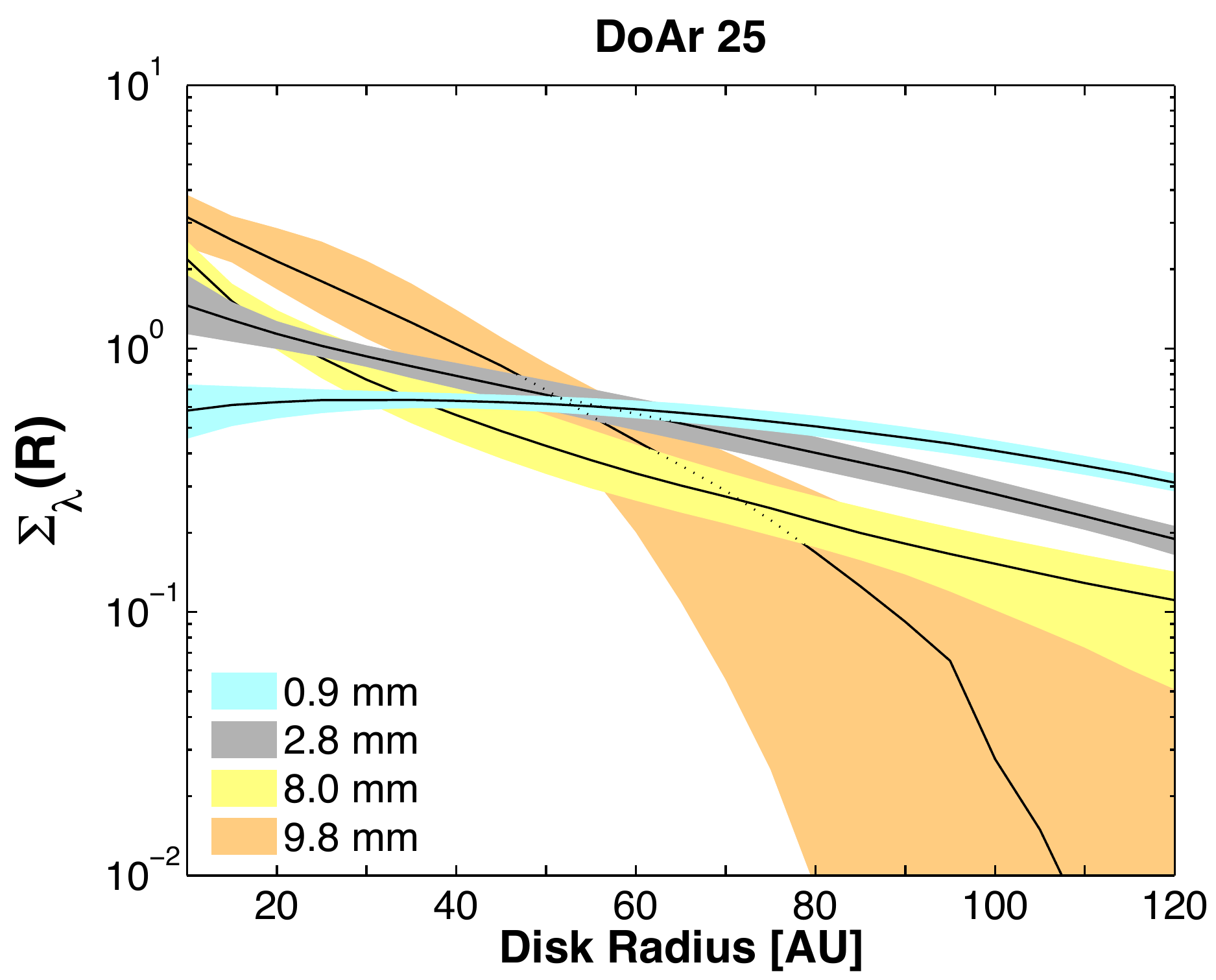}
\includegraphics[scale=0.31]{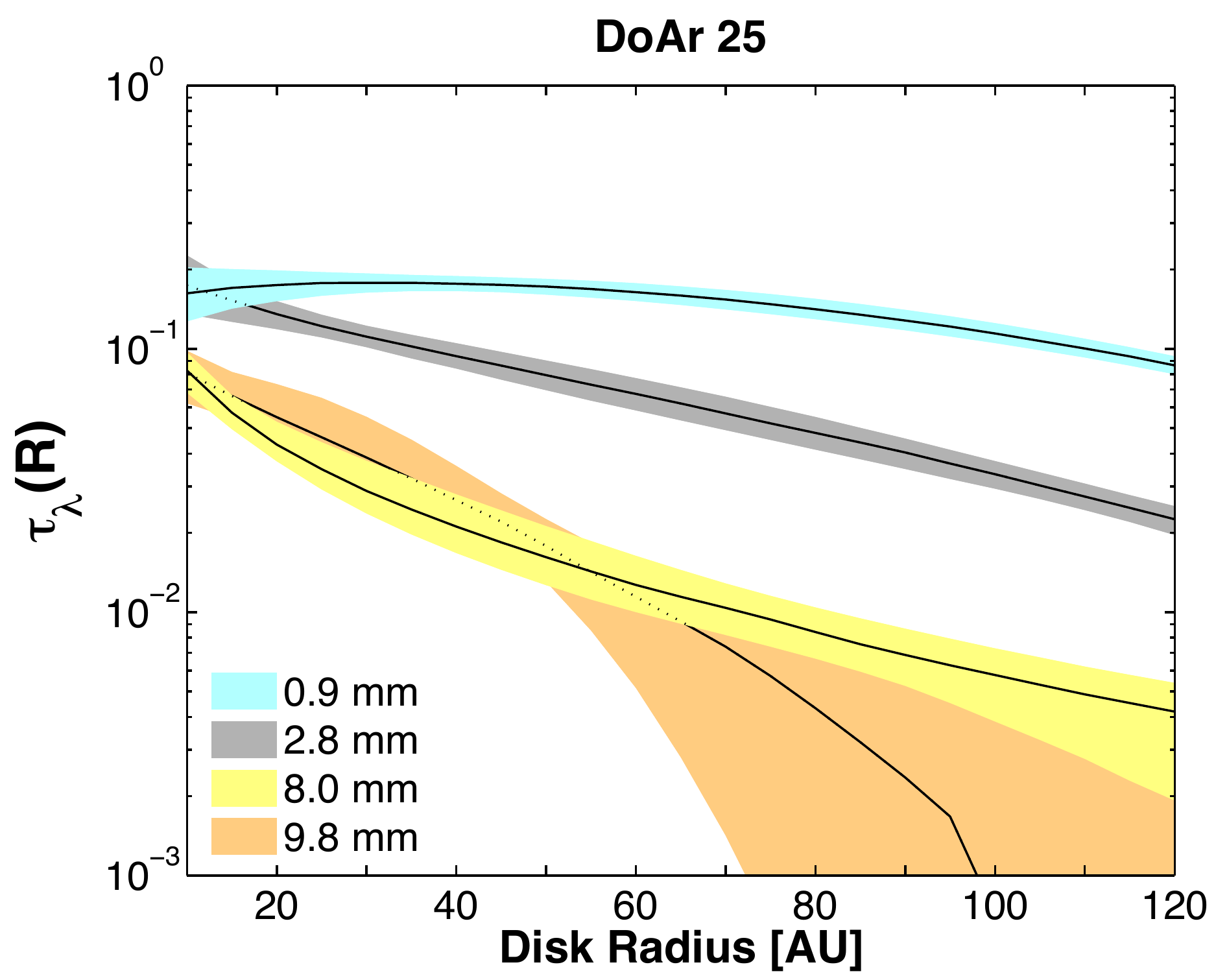}
\caption{\emph{Left and middle:} $T_{\lambda}(R)$ and $\Sigma_{\lambda}(R)$ inferred from separate modeling of our multi-wavelength observations of \doardisk, assuming a constant dust opacity with radius. \emph{Right:} Optical depth $\tau_{\lambda}(R)=\kappa_{\lambda}\times\Sigma_{\lambda}(R)$ inferred from separate modeling of \doardisk\ multi-wavelength observations, assuming a radially constant $\kappa_{\lambda}$.
Colored regions: $3\sigma$ confidence interval constrained by our observation; continuous line: best-fit model; dashed line on left panel: mean temperature profile.
The different $\Sigma(R)$ and $T(R)$ profiles for each wavelength imply a varying dust opacity with radius, and because of this, none of them is the true surface density and temperature profile of the disk.}\label{DoAr25_structure}
\end{center}
\end{figure*}

From each separate model fitting we characterize the disk surface density and temperature profile ($\Sigma(R)$ and $T(R)$) at each wavelength. Figures \ref{CYTau_structure} and \ref{DoAr25_structure} present the best-fit and $3\sigma$ constraints on the disk temperature, surface density, and optical depth ($\tau_{\lambda} = \kappa_{\lambda}\Sigma(R)$), obtained from modeling each wavelength independently. 
The temperature profiles inferred from observations at different wavelengths are very similar (this is expected, since dust of different sizes in the disk should have reached radiative equilibrium and be at the same temperature). In particular, the best-fit temperature at each wavelength differs from the mean temperature profile by at most $\sim3$~K for \cytau\ and $\sim5$~K for \doardisk. 
However, the surface density profiles that were inferred separately from each observation (middle panels of Figures \ref{CYTau_structure} and \ref{DoAr25_structure}), are different at each observed wavelength. This is clearly not physical (the dust emission arises from the same disk at all wavelengths!), and cannot be reconciled through a global change in the dust opacity. We conclude that we need to consider a change in the dust opacity as a function of location (specifically, radius) in the disk, in order to explain the observed dust emission. This is explored further in \S\ref{beta_constraints} below. Finally, the right panels of Figures \ref{CYTau_structure} and \ref{DoAr25_structure} show the inferred optical depth, demonstrating that the dust emission is optically thin at all wavelengths for $R>15$~AU (well within our spatial resolution at all wavelengths for both disks). Furthermore, the emission would be optically thin even if the mid-plane dust temperature were lower by a factor of a few.

\subsection{Radial variations of the dust opacity}
\label{beta_constraints}

Since the dust emission is in the optically thin regime for both \cytau\ and \doardisk\ (right panel of Figures~\ref{CYTau_structure} and \ref{DoAr25_structure}), the observed emission will directly depend on three physical quantities: the dust opacity, the dust mass surface density, and the temperature, such that $S_{\nu} \propto \kappa_{\lambda}  \Sigma  B_{\nu}(T)$. Thus, our observations provide a constraint for the product $\kappa_{\lambda} \times \Sigma \times B_{\nu}(T)$, where the opacity, surface density, and temperature may vary with radius, but only the dust opacity is a function of the observed wavelength. Our model fitting presented above -- which assumes a constant dust opacity with radius and where each wavelength is fitted separately -- results in a wavelength-dependent $\Sigma(R)$, signifying that the assumption of a radially constant dust opacity is not warranted by the data. To reconcile these wavelength-dependent $\Sigma_{\lambda}(R)$ and $T_{\lambda}(R)$ (Figures \ref{CYTau_structure} and \ref{DoAr25_structure}), as well as the wavelength-dependent disk structure seen in the normalized visibility profiles (Figures~\ref{vis_profile_cytau} and \ref{vis_profile_doar25}) we require a change in the dust properties as a function of radial distance from the central star, for both the \cytau\ and \doardisk\ disks.

We focus on the interpretation that the spectral index of the dust opacity, $\beta$, is changing with radius, which we obtain following the procedure described by \citet{2012Perez} and \citet{2010Isella}, and summarize here. Since the constraints found in Section~\ref{sigma_temp_results} reproduce the data well, these must be consistent with the \emph{true} (but unknown) disk structure:
\begin{equation}
\kappa_{\lambda} \: \Sigma_{\lambda}(R) \: B_{\nu}(T_{\lambda}(R)) = \overline \kappa_{\lambda}(R) \: \overline\Sigma(R) \: B_{\nu}(\overline T(R))
\label{kappa_equality}
\end{equation}
where $\overline\kappa_{\lambda}(R)$, $\overline\Sigma(R)$, and $\overline T(R)$ are the true disk physical quantities, and the left-hand side of this equation corresponds to our multi-wavelength constraints for the disk structure. Thus, two different wavelengths, $\lambda_1$ and $\lambda_2$ (and corresponding frequencies, $\nu_1$ and $\nu_2$), can be related by
\begin{eqnarray}
\left( \frac{\lambda_1}{\lambda_2} \right)^{-\beta(R)} &=& \left( \frac{\lambda_1}{\lambda_2} \right)^{-\beta_C} 
\times \frac{\Sigma_{\lambda_1}(R) \frac{B_{\nu_1}(T_{\lambda_1}(R))}{B_{\nu_1}(\overline T(R))}}{\Sigma_{\lambda_2}(R) \frac{B_{\nu_2}(T_{\lambda_2}(R))}{B_{\nu_2}(\overline T(R))}}
\label{beta1}
\end{eqnarray}
assuming that at long wavelengths the dust opacity follows a power-law behavior ($\kappa_{\lambda} \propto \lambda^{-\beta}$), and $\beta_C$ corresponds to the spectral slope of the assumed constant dust opacity with radius, derived from the integrated SED used in the initial models. 
A useful prescription to infer radial variations of the opacity spectral slope ($\Delta \beta(R) = \beta(R) - \beta_C$) in dual-wavelength observations is derived from this last equation (see also equation (9) from \citet{2010Isella}):
\begin{equation}
\Delta \beta(R) = - \log^{-1} \left[ \frac{\lambda_1}{\lambda_2} \right] \times \log \left[ \frac{\Sigma_{\lambda_1} B_{\nu_1}(T_{\lambda_1})/B_{\nu_1}(\overline T)}{\Sigma_{\lambda_2} B_{\nu_2}(T_{\lambda_2})/B_{\nu_2}(\overline T)} \right]
\label{beta2}
\end{equation}
where the dust surface density and temperature depend on the distance from the star $R$.
Equation \ref{beta2} implies that in logarithmic space $\Delta \beta(R)$ is the slope of a straight line at points \{$x = \log(\lambda), y = \log(\Sigma_{\lambda} B_{\nu}(T_{\lambda})) / B_{\nu}(\overline T)$\}.

Note that if the emission is in the Rayleigh-Jeans (R-J) domain: $B_{\nu}(T)\propto T$, and thus the true dust temperature ($\overline T(R)$) will not be needed to derive $\Delta \beta(R)$. However, at cold temperatures in the outer disk and/or at short wavelengths, the R-J assumption breaks down and we require an estimate of $\overline T(R)$. As mentioned in Section~\ref{sigma_temp_results}, the temperature constraints from these multi-wavelength data are quite similar, thus we assume the true dust temperature to be the mean $T_{\lambda}(R)$, shown as a dashed line on the left panels of Figures \ref{CYTau_structure} and \ref{DoAr25_structure}. Note that a change in the estimate of $\overline T(R)$ will have a minimal impact in the $\Delta \beta(R)$ constraint, as $\overline T(R)$ only appears in the ratio of Planck functions evaluated at two different wavelengths in equation~\ref{beta2} (and close to the R-J limit the dependence with $\overline T(R)$ is negligible). Nevertheless, we tested two different assumptions where $\overline T(R)$ was equal to \emph{twice} and \emph{half} the mean of $T_{\lambda}(R)$. These substantial changes of $\times2$ only changed our constraints in $\Delta\beta(R)$ by $\sim10$\% or less throughout the disk.

\begin{figure}
\begin{center}
\includegraphics[scale=0.42]{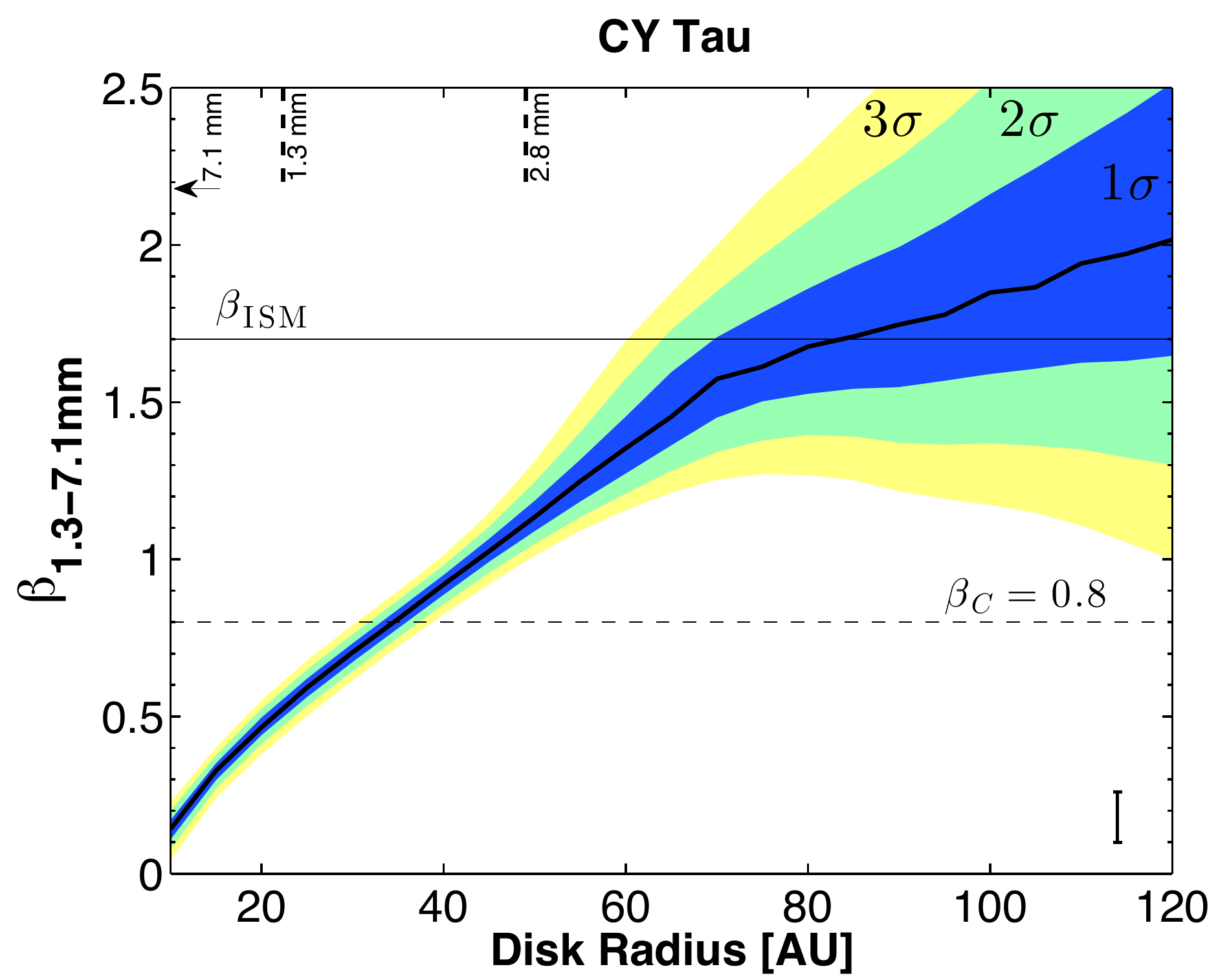}
\caption{Spectral index of the dust opacity ($\beta$) as a function of radius, inferred from multi-wavelength observations of the \cytau\ disk. Black line: best-fit $\beta(R)$; colored areas: confidence interval constrained by our observations. The vertical dashed-lines represent the smallest spatial scale probed at each wavelength, the arrow indicates that at 7.1~mm the smallest spatial scale is outside of the range plotted in this figure, corresponding to 42~mas or 6~AU. The errorbar at the bottom right indicates the systematic uncertainty on $\beta(R)$ resulting from absolute flux scale uncertainty.}
\label{beta_CYTau}
\end{center}
\end{figure}

\begin{figure}
\begin{center}
\includegraphics[scale=0.42]{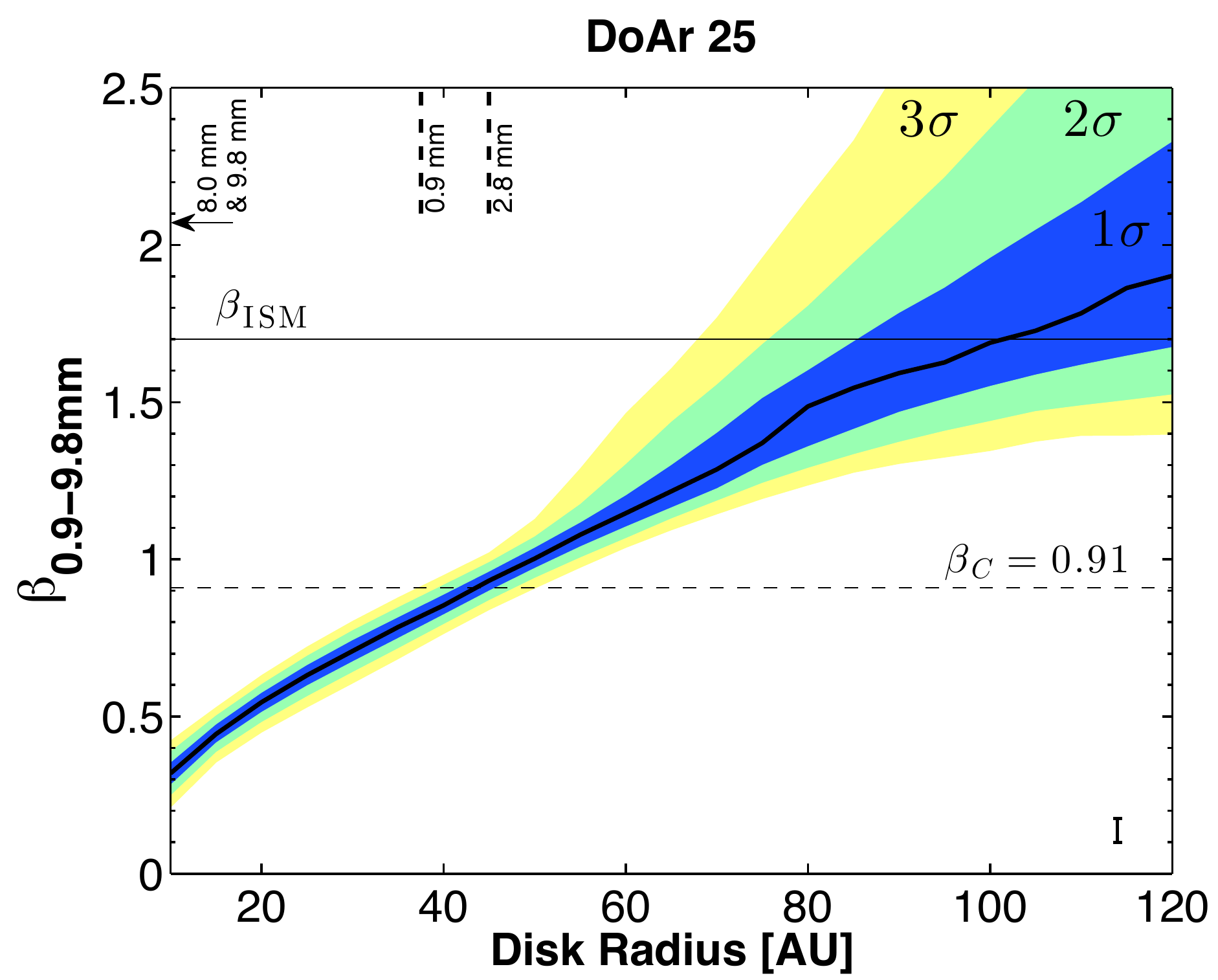}
\caption{Spectral index of the dust opacity ($\beta$) as a function of radius, inferred from multi-wavelength observations of the \doardisk\ disk. Black line: best-fit $\beta(R)$; colored areas: confidence interval constrained by our observations. The vertical dashed-lines represent the smallest spatial scale probed at each wavelength, the arrow indicates that at 8.0 and 9.8~mm the smallest spatial scale is outside of the range plotted in this figure, corresponding to 48~mas or 6~AU at 8.0~mm, and 58~mas or 7~AU at 9.8~mm. The errorbar at the bottom right indicates the systematic uncertainty on $\beta(R)$ resulting from absolute flux scale uncertainty.}
\label{beta_DoAr25}
\end{center}
\end{figure}

Since the result of the MCMC algorithm is a sampled posterior PDF for the parameter space, we can construct a PDF for the product $\Sigma_{\lambda}\times B_{\nu}(T_{\lambda})/ B_{\nu}(\overline T)$ at each radius and for each wavelength that has been modeled. 
To determine $\Delta \beta(R)$ we use a Monte Carlo simulation, where the PDF of $\Sigma_{\lambda} \times B_{\nu}(T_{\lambda})/ B_{\nu}(\overline T)$ is randomly sampled at each wavelength; a random sampling of
$\Delta \beta(R)$ will thus correspond to the slope of the line that intersects the points \{$x = \log(\lambda)$, $y = \log(\Sigma_{\lambda} B_{\nu}(T_{\lambda}))/B_{\nu}(\overline T)$\}. 
By obtaining a large number of random samples of $\Sigma_{\lambda} \times B_{\nu}(T_{\lambda})/ B_{\nu}(\overline T)$ we populate the PDF of $\Delta\beta(R)$ at each radius. 
We find the best-fit value of $\Delta\beta$ at a radius $R$ from the peak of this PDF, and we derive confidence intervals from the region of the distribution that contains 68.3\%, 95.5\%, and 99.7\% of all samples at equal probability (as described in Section~\ref{modeling_emission} for the uncertainties of the model parameters).
These steps are performed for all radii sampled by the data, in order to obtain a  best-fit value and inferred constraints on $\beta(R) = \beta_C + \Delta\beta(R)$ for \cytau\ and \doardisk.

Figures~\ref{beta_CYTau} and \ref{beta_DoAr25} present the observational constraint on $\beta$ as a function of radius, obtained for the circumstellar disks of \cytau\ and \doardisk. 
The values of $\beta$ allowed by these multi-wavelength observations are significantly different from (and below) the ISM value of the dust opacity slope ($\beta_{\rm ISM}\sim1.7$) for $R\lesssim60$~AU in the case of \cytau, and $R\lesssim70$~AU for \doardisk. 
Furthermore, a gradient on $\beta$ with radius is found in both disks. This gradient can only be consistent with a constant value of $\beta$ when the uncertainties in $\beta(R)$ are extended to $7\sigma$.
We note that the uncertainty in the absolute flux density scale will potentially introduce an additional systematic offset in the constrained values of $\beta(R)$. But because a fractional uncertainty in the flux density scale affects all radii equally at a particular wavelength, the overall shape and significance of the deviation in $\beta(R)$ from a constant value are {\it not} affected by uncertainties in the flux density scale. Given the 10-15\% uncertainty in the absolute flux density scales for the three telescopes used, we estimate that the level of the systematic offset introduced to be $\sim 0.16$ for \cytau\ and $\sim 0.08$ for \doardisk, indicated by a vertical errorbar in the bottom-right corner of Figures~\ref{beta_CYTau} and \ref{beta_DoAr25}. This systematic uncertainty is smallest for \doardisk, because observations of this disk were obtained at wavelengths that are further separated than for \cytau, reducing the uncertainty due to the increase in the lever arm used to infer $\beta$.

\subsection{Radial variation of the dust opacity under the assumption of grain growth}

As discussed in the introduction, the inferred changes in the dust opacity for \cytau\ and \doardisk\ may arise from changes in grain size, composition, or grain geometry. \citet{2006Draine} explored the frequency dependence of the dust opacity for different materials which may have enhanced emissivity at long wavelengths and thus can create low values of $\beta$ with small grains. He concluded that changes in composition are unlikely to explain $\beta<1$ and advocates for changes in the grain-size distribution to explain the common finding of $\beta\lesssim1$ in unresolved observations of protoplanetary disks (for references see \S1). On the other hand, changes in grain porosity can produce low values of $\beta$ (see, e.g., Figure 11 of \citet{2014Kataoka}). However, the low filling factor needed to reach $\beta<1$ requires quite large fluffy grains (e.g., larger than meter-size for filling factors of $10^{-1}-10^{-2}$). Given the complexity of exploring changes in composition, geometry, and size of dust grains simultaneously, and given that the option of large grains can more simply reproduce the observed changes in $\kappa_{\lambda}$, we assume that the radial variations of the dust opacity (observed as changes in $\beta$ in the previous section) are only caused by changes in $a_{max}$ with radius. We now estimate the range of allowed maximum grain sizes at each location of the disk for \cytau\ and \doardisk. 

As discussed by \citet{2012Perez}, the determination of $\beta(R)$ depends on the assumption that the dust opacity follows a power-law with wavelength: $\kappa_{\lambda} \propto \lambda^{-\beta}$. This essential assumption might not be the case, as non-power-law dependence with wavelength in $\kappa_{\lambda}$ can be seen on dust populations which are limited to a maximum grain size ($a_{max}$) of sub-millimeter and mm-sized particles \citep[see e.g., ][]{2006Draine}. Thus, variations in the dust properties of these disks, particularly $a_{max}$, should be directly constrained without this assumption.

Following the method outlined by \citet{2012Perez}, we find the best-fit dust opacity (which depends on $a_{max}$), along with the best-fit surface density, that is needed to reproduce the constraints on $\kappa_{\lambda}\times\Sigma_{\lambda}\times B_{\nu}(T_{\lambda})$ found in Section~\ref{sigma_temp_results}.
Since the temperature profile at each wavelength ($T_{\lambda}$) is found to be similar at all wavelengths, we assume the dust temperature is equal to the mean $T_{\lambda}$ at each radius (see dashed line on left panel of Figures \ref{CYTau_structure} and \ref{DoAr25_structure}).
At this point, we include the uncertainty in the absolute flux density scale in our modeling, which ranges between 10-15\% for the three telescopes used. 
This improvement to our method is important, as it introduces another source of error in the constraints obtained from the flux density at each wavelength, and impacts the error budget in the characterization of radial variations of the maximum grain size. 
We note that the uncertainty in the flux density scale as a function of wavelength was not included in our previous analysis of AS~209 \citep[e.g., ][]{2012Perez}. 
For each MCMC realization that reproduces the disk surface brightness as a function of radius, we draw a random number from a normal distribution centered at 1.0 and whose standard deviation ($\sigma$) corresponds to the flux scale uncertainty. Depending on the wavelength, this would correspond to $\sigma = 0.1$ or  $\sigma = 0.15$.
We then scale the surface density normalization of each MCMC realization by a different random number, ensuring that the uncertainty in the absolute flux density scale is reflected in the resulting probability distribution.

\begin{figure*}
\begin{center}
\includegraphics[scale=0.42]{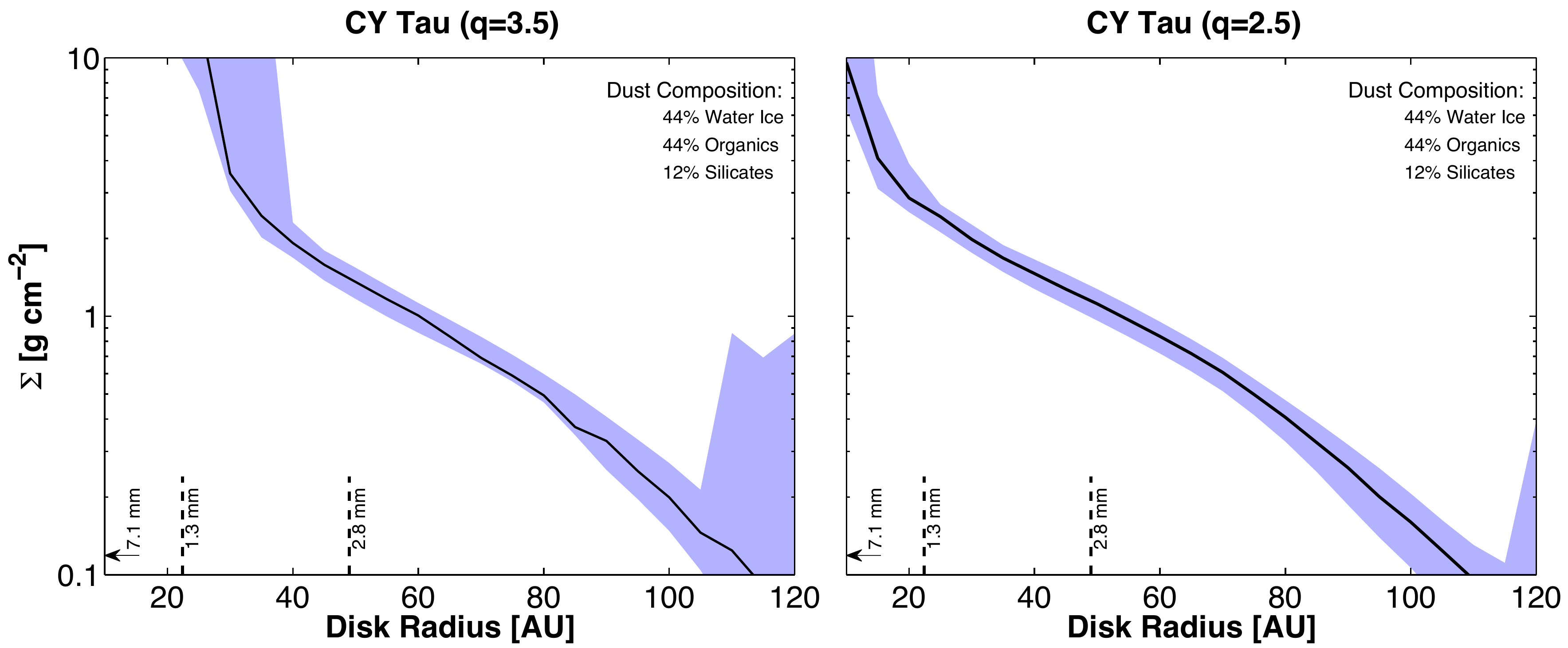}
\includegraphics[scale=0.42]{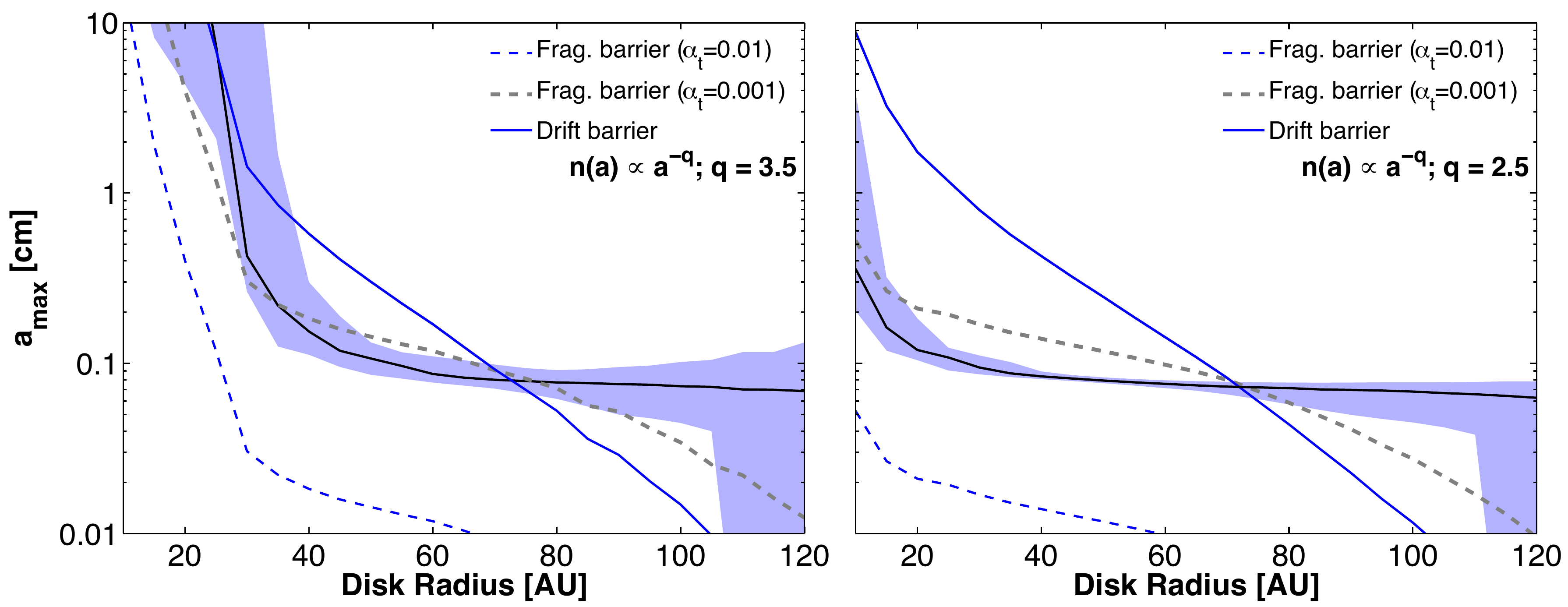}
\caption{Constraints for the \cytau\ disk on the surface density (\emph{top panels}) and maximum grain size (\emph{bottom panels}), for two grain size distributions $q=3.5$ (\emph{left panels}), $q=2.5$ (\emph{right panels}). 
The best-fit $\Sigma(R)$ (top) and $a_{max}(R)$ (bottom) are indicated by a continuous black line, while the shaded region represents our confidence interval at $3\sigma$. We compare our observational constraints with theoretical grain evolution models by \citet{2012Birnstiel} that include fragmentation (dashed lines, bottom panels), or radial drift (blue continuous line, bottom panels). For the fragmentation barrier we explore two different levels of turbulence in the \cytau\ disk: $\alpha_t=0.01$ (blue dashed line) and $\alpha_t=0.001$ (grey dashed line), with $v_{\mathrm{frag}}=10$~m~s$^{-1}$ in both cases. The smallest spatial resolution probed at each wavelength is indicated by a vertical dashed line in the top panels (at 7.1~mm, the smallest spatial scale corresponds to 42~mas or 6~AU, outside of the range plotted in this figure and indicated by an arrow).}
\label{amax_cytau}
\end{center}
\end{figure*}

\begin{figure*}
\begin{center}
\includegraphics[scale=0.42]{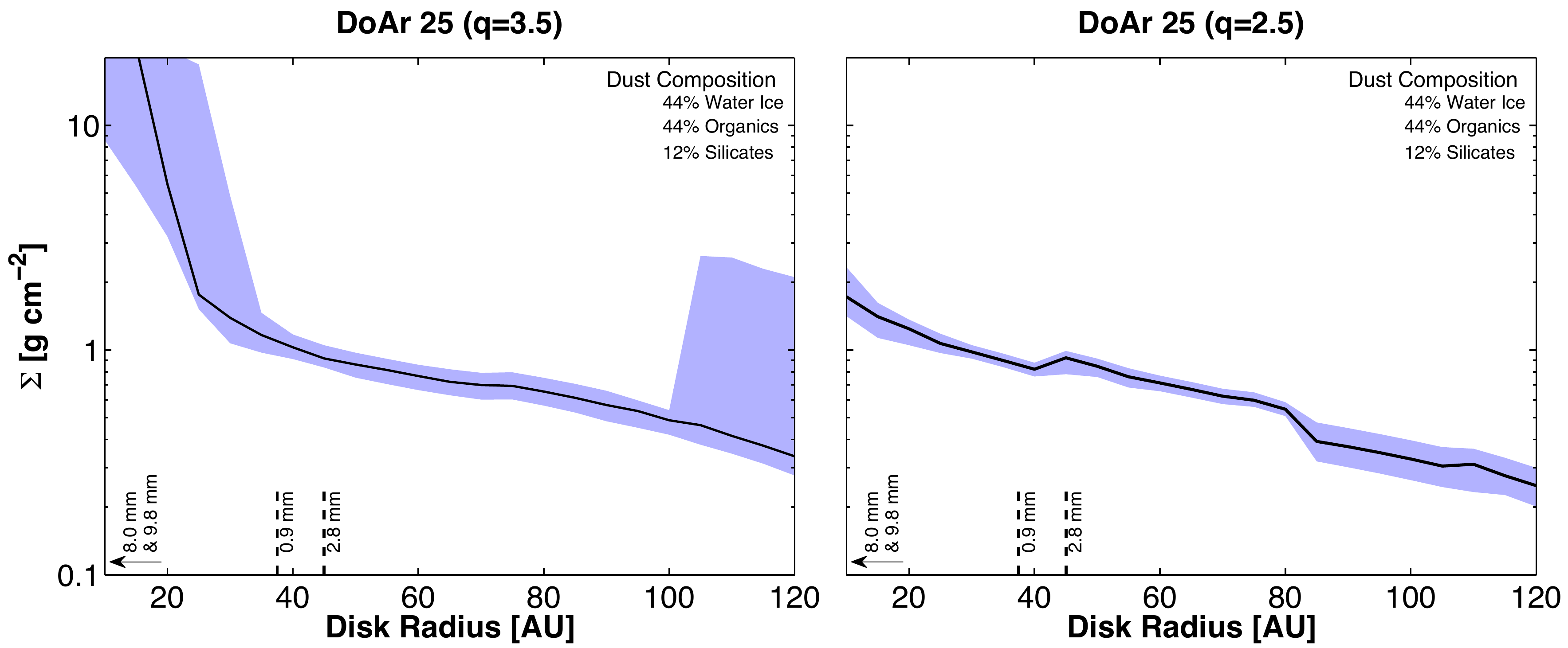}
\includegraphics[scale=0.42]{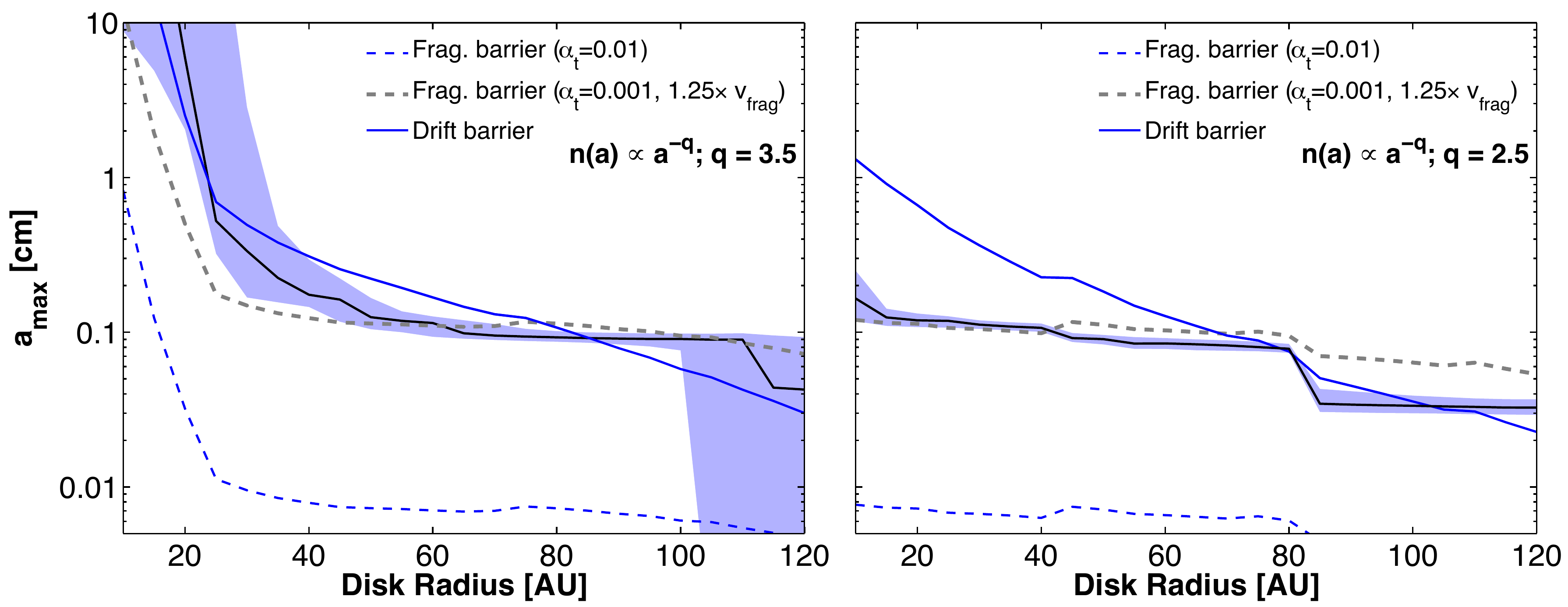}
\caption{Constraints for the \doardisk\ disk on the surface density (\emph{top panels}) and maximum grain size (\emph{bottom panels}), for two grain size distributions $q=3.5$ (\emph{left panels}), $q=2.5$ (\emph{right panels}). 
The best-fit $\Sigma(R)$ (top) and $a_{max}(R)$ (bottom) are indicated by a continuous black line, while the shaded region represents our confidence interval at $3\sigma$. We compare our observational constraints with theoretical grain evolution models by \citet{2012Birnstiel} that include fragmentation (dashed lines, bottom panels), or drift (blue continuous line, bottom panels). For the fragmentation barrier we explore two different levels of turbulence and fragmentation velocity in the \doardisk\ disk: $\alpha_t=0.01$ with $v_{\mathrm{frag}}=10$~m~s$^{-1}$ (blue dashed line), and $\alpha_t=0.001$ with $v_{\mathrm{frag}}=12.5$~m~s$^{-1}$ (grey dashed line). The spatial resolution at each wavelength is indicated by a vertical dashed line in the top panels (FWHM for 8.0 and 9.8~mm, HWHM for 2.8 and 0.9~mm).}
\label{amax_doar25}
\end{center}
\end{figure*}

Assuming the same dust composition as in Section~\ref{modeling_emission} we place $3\sigma$ confidence limits for $a_{max}(R)$ and $\overline\Sigma(R)$, while exploring two representative grain size distributions: a \emph{steep} distribution with $q=3.5$ and a 
\emph{shallow} distribution with $q=2.5$. The chosen distributions are motivated by the resulting size-distribution of a collisional cascade \citep{1969JGR....74.2531D} and the measured grain size distribution in the ISM \citep{1977ApJ...217..425M}, which are both characterized by $n(a)\propto a^{-3.5}$, while a less steep distribution of $q=2.5$ is selected to observe the effects of a reduced total number of smaller particles.
These results are presented as the shaded regions in Figures~\ref{amax_cytau} and \ref{amax_doar25} for \cytau\ and \doardisk, respectively.
We find that within $\sim100$~AU from the central protostar, grains have grown up to at least up to $\sim0.4$~mm for \cytau\ and $\sim0.2$~mm for \doardisk.
The best-fit $a_{max}(R)$ displays a gradient with radius, where smaller grains are present in the outer disk and larger grains are present in the inner disk. 
However, the rise of $a_{max}$ towards smaller radii strongly depends on the assumed slope of the grain size distribution.
This occurs because a \emph{shallow} grain-size distribution  will have more large grains (by number) than a \emph{steep} distribution. 
Since large grains dominate the emission at these wavelengths, we can reproduce the observed emission with a maximum grain size that can be smaller for a shallow distribution  than for a steep distribution.
This explains the difference in our $a_{max}$ constraints for the representative slopes $q=2.5$ and $q=3.5$ that we selected. Unfortunately, current observations do not have the necessary sensitivity to discern the value of the grain size distribution slope.

\section{Discussion}
\label{discussion}

\subsection{Comparison with theory}

\cite{2012Birnstiel} presented simple analytical forms for the variation of $a_{max}$ with radius for two different regimes: when a population of dust grains is limited by fragmentation of dust particles, whose differential velocity is driven by turbulence in the disk:
\begin{equation}
a_{max}^{\rm frag}(R) \sim \frac{2}{3\pi} \frac{\Sigma_{gas}(R)}{\rho \alpha_t} \frac{v^2_{\rm frag}}{c^2_s(R)}
\label{frag_barrier}
\end{equation}
and when a population of dust grains is limited by radial drift, as larger particles are removed from the population and accreted onto the star due to gas drag:
\begin{equation}
a_{max}^{\rm drift}(R) \sim \frac{2}{\pi} \frac{\Sigma_{dust}(R)}{\rho} \frac{v^2_{\rm K}(R)}{c^2_s(R)} \left|\frac{d\ln P}{d\ln R}\right|^{-1}.
\label{drift_barrier}
\end{equation}
Here $\Sigma_{gas}$ and $\Sigma_{dust}$ are the gas and dust surface density profile, $\rho$ is the dust grain internal density, $\alpha_t$ is the disk turbulence parameter as presented by \citet{1973Shakura}, $v_{\rm frag}$ is the velocity at which a collision between dust grains will result in fragmentation rather than growth, $v_{\rm K}$ is the local Keplerian velocity, and $c_s$ is the local sound speed of the gas.

Employing our constraints on the disk surface density, which assume a constant gas-to-dust ratio of 100:1 over the entire disk, we compared our derived observational constraints on $a_{max}$ vs. radius with these grain growth models, adopting standard values for the turbulence ($\alpha_t = 0.01$), dust density ($\rho=1.3$ g cm$^{-3}$, from the \citet{1994Pollack} abundances adopted in this paper), and fragmentation velocity \citep[$v_{\rm frag} = 10$~m s$^{-1}$][]{2012Birnstiel}.
Also, we verify that mm-sized particles are in the Epstein regime throughout the disk (with a Stokes number $\rm St<1$), which is a necessary condition for the use of equation~\ref{drift_barrier} as the size limit in radial drift.
For this particular set of turbulence and fragmentation parameters, the theoretical $a_{max}$ curves for \cytau\ and \doardisk\ are shown in Figures~\ref{amax_cytau} and \ref{amax_doar25}, respectively, where the continuous line illustrates the maximum grain size for a drift-dominated population, and the dashed blue line correspond to a fragmentation-dominated population.
Note that since our constraint on the surface density with radius depends on the adopted grain size distribution slope, $q$, the drift and fragmentation barriers will depend on $q$ as well.

In both disks and with this standard set of parameters, the fragmentation barrier under-predicts the maximum grain size achievable.
Following equation~\ref{frag_barrier}, a better correspondence could be found if: (1)~the disk turbulence were lowered to $\alpha_t\sim10^{-3}-10^{-4}$, or (2)~the gas-to-dust ratio were higher than 100:1 by an order of magnitude, or (3)~the critical velocity for fragmentation were higher than $v_{\rm frag} = 10$~m s$^{-1}$. 
This last option seems unlikely if grains are compact, given that threshold velocities for fragmentation and/or bouncing of up to a few m s$^{-1}$ have been measured experimentally for compact centimeter and decimeter-sized grains (see e.g., \citet{2013ApJ...769..151D} for silicate grains, \citet{2010Icar..206..424H} and \citet{2015A&A...573A..49H} for icy grains, and \citet{2008Blum_Wurm} for a review).
On the other hand, if grains are porous \citep[e.g. ][]{2014Kataoka}, fragmentation could occur at higher threshold velocities. However, porous grains have larger cross sections resulting in mass loss through collisional erosion rather than fragmentation, which similarly halts the growth of these grains \citep{2015Krijt}.
Now, the second option of a higher gas-to-dust ratio needs to be pursued with measurements of the total gas mass in these disks in particular, but this alternative also seems unlikely given the recent results in the Taurus-Auriga star forming region where low gas-to-dust ratios (well below the ISM standard of 100:1) are inferred for a large number of circumstellar disks by \citet{2014ApJ...788...59W}.
Finally, the first option of a quiescent disk of low turbulence ($\alpha_t < 0.01$) for the disks around \cytau and \doardisk (and also AS~209) seems plausible, and would further augment grain growth. 
We test this option by calculating the expected maximum grain size when the disk turbulence is an order of magnitude lower ($\alpha_t = 0.001$), and we find a better correspondence with our $a_{max}$ constraints on \cytau\ and \doardisk\ (for this last disk we also increased the fragmentation velocity by 25\%).
Note that values of $\alpha_t\sim10^{-3}-10^{-4}$ would still produce turbulent relative velocities high enough for fragmentation, a condition attained when $\alpha_t > \frac{1}{3}(\frac{v_{\rm frag}}{c_s})^2$, and confirmed for both \cytau\ and \doardisk.
The prediction of low turbulence should be tested with future observations of the turbulent linewidth in these circumstellar disks.

At the same time, the radial drift barrier roughly agrees with our $a_{max}$ inference in both disks. 
For \doardisk, the growth barrier imposed by a radial drift of macroscopic particles is within a factor of a few of the observational constraint, and the same occurs for \cytau\ at least up to $80$~AU. 
Most likely it is both mechanisms that are at play \emph{simultaneously} in these disks.
Finally, we note that at a radii $\gtrsim100$~AU, the population of grains in \cytau\ is above both barriers by a factor of few at least, and this is the case for both a shallow or a steep grain size distributions.
This result is readily appreciated in the \cytau\ data, where millimeter-wave emission at 1.3 and 2.8~mm is observed at large distances from the star, indicating the presence of mm-sized grains at $\sim100$~AU.
If the gas surface density distribution were different from the dust surface density we measured, the pressure gradient term $\left|\frac{d\ln P}{d\ln R}\right|$ in equation~\ref{drift_barrier} would change, and thus a region of higher gas pressure, i.e., a local maximum in $P(R)$, could trap mm-sized dust grains in the outer disk of \cytau. In such a scenario, grains can overcome the radial drift barrier that is expected when the gas distribution just monotonically increases with radius \citep[e.g.,][]{2012A&A...538A.114P}. 
The origin of these pressure enhancements is strongly debated: locally they may arise from turbulent eddies \citep{1997Icar..128..213K,2005ApJ...634.1353J}, while on a global disk scale they may emerge from Rossby-wave instabilities \citep{1999ApJ...513..805L,2000ApJ...533.1023L}, long-lived vortices \citep{1995A&A...295L...1B}, or from the gap opened by one or multiple planets \citep{2012A&A...545A..81P}.
However, higher angular resolution observations of the distribution of small and large grains, as well as observations of tracers in the gas emission, are needed to understand what is the likely cause of these large grains still present in the outer disk of \cytau.

\begin{figure*}
\begin{center}
\includegraphics[scale=0.42]{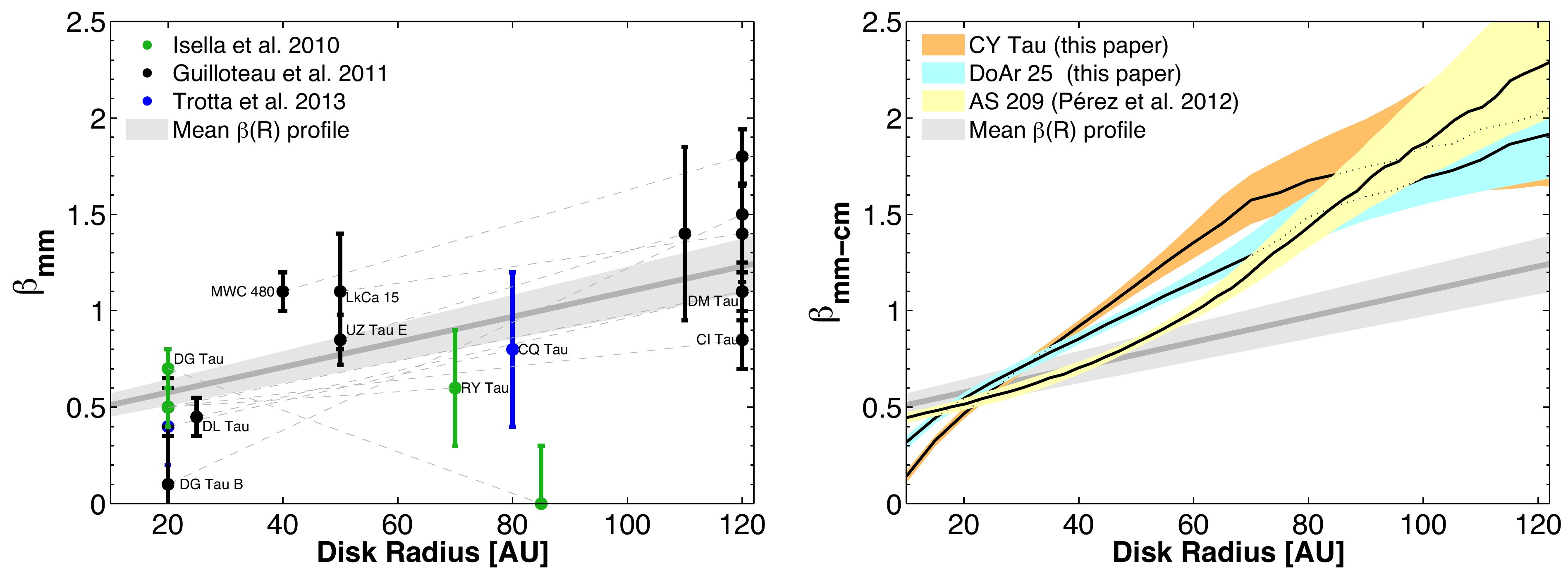}
\caption{ Left: Observational constraints on the spectral index of the dust opacity $\beta$, with $1\sigma$ error bars, as measured from published resolved millimeter observations, where all but CQ~Tau have been observed only at 1 and 3~mm \citep{2010Isella,2011Guilloteau,2013Trotta}. Note that the majority of the previously available constraints are still consistent with a constant value of $\beta$ across the disk at $3\sigma$. The mean profile of $\beta(R)$ based on these observations is shown by the gray shaded region. Right: The radial variation of $\beta$ for the three circumstellar disks analyzed in a consistent fashion: \cytau\ and \doardisk\ (this paper), and AS~209 \citep{2012Perez}, where the shaded regions show the $1\sigma$ constraints. We note that at the $3\sigma$ level the 3 disks analyzed here at consistent with each other and with the average $\beta(R)$ profile from the literature.}
\label{comparison_beta}
\end{center}
\end{figure*}

\subsection{Comparison with other disks}

Over the past several years there have been multiple examples of circumstellar disks where the interpretation of radial variations of the dust properties is supported by resolved observations at mm and cm wavelengths \citep{2010Isella,2011Guilloteau,2013Trotta,2014Menu,2015ApJ...808..102K}.
For example, in TW~Hya, \citet{2014Menu} found that a radially homogeneous distribution of small and large grains (larger than $\sim 100$~$\mu$m) was not a suitable representation of the observations, and a steeper $\Sigma(R)$ for the larger grains was required to fit these multi-wavelength dataset.
And recently \citet{2015ApJ...808..102K} studied a sample of disks and found that those with a steep midplane density gradient generally possess a smaller value of $\beta$, indicative of dust growth and radial evolution.
The largest sample to date has been gathered by \citet{2011Guilloteau}, where dual-wavelength observations of disks in Taurus-Auriga at angular resolutions between $0.4''-1.0''$ were presented. 
This study found evidence for changes in the value of $\beta$ for several sources including \cytau.
Albeit limited by the angular resolution of the lowest frequency band, which corresponds to $\sim100$~AU resolution for their 2.7~mm observations of \cytau, these authors infer a change from $\beta\sim0$ in the inner disk to high $\beta$-values in the outer disk of this young star. 
However, given the closeness in frequency of the two wavelengths studied by \citet{2011Guilloteau}, together with the lack of high spatial resolution at 2.7 mm, implies that their constraint is consistent with a constant value of $\beta$ at the $3\sigma$ level for this particular object. 
As shown in Section~\ref{beta_constraints}, our increased wavelength coverage and high sensitivity allows us to discriminate radial variations of the dust emissivity index for \cytau\ with high significance, where a constant value of $\beta$ is excluded at more than $7\sigma$. This arises because the uncertainty in the spectral index ($\sigma_{\alpha}$) is directly related to the wavelength coverage and SNR of the observations. In the case of dual-wavelength observations, simple error propagation results in an uncertainty of $\sigma_{\alpha}^2 = (\ln(\nu_1/\nu_2))^{-2} (\mathrm{SNR}_{\nu_1}^{-2} + \mathrm{SNR}_{\nu_2}^{-2} )$, which indicates that an efficient way to reduce the uncertainty in $\alpha$, and thus in $\beta$, is to expand the wavelength coverage.

To compare radial changes in the dust properties for different circumstellar disks, we have gathered measurements from the literature that attempt to constrain $\beta(R)$, mostly from observations at the 1 and 3~mm bands. Since most of the available measurements barely resolve each circumstellar disk, we employ only two representative measurements of $\beta$, one from the inner disk regions and another from the outer disk. Thus, for each published result we compile the value of $\beta$ and its $1\sigma$ uncertainty at the best resolution of the observations (between $\sim20-50$ AU depending on each target) to illustrate the inner disk constraint in $\beta$, and at either the observed outer disk radius or at $R=120$~AU, whichever is smallest, to represent the constraint of $\beta$ in the outer disk (except for RY~Tau and DG~Tau, which at the outer disk radius have an unconstrained value of $\beta$; for these we select the outermost radii for which a $\beta$ constraint is still significant).

The compiled literature measurements are presented in the left panel of Figure~\ref{comparison_beta}, where we have computed the slope in $\beta(R)$ between the inner and outer disk for each star (dashed lines), and the average slope in $\beta(R)$ for all the disks analyzed in the literature (grey shaded region). The steepest change in the dust emissivity spectral index occurs for DG~Tau~B \citep[where $\beta = 0.1\pm0.3$ at 20AU, increasing to $\beta = 1.5\pm0.3$ at 120 AU;][]{2011Guilloteau}. 
Excluding the added uncertainty in $\beta$ due to the absolute flux scale uncertainty for the two millimeter-wave bands observed for DG~Tau~B, the resulting change in $\beta(R)$ for this object is comparable to the constraints found for \cytau, \doardisk, and AS~209, which are presented together on the left panel of Figure~\ref{comparison_beta}. 

Interestingly, the mean $\beta(R)$ profile from literature measurements is less steep than for the 3 disks analyzed in a consistent fashion. This could be due to the fact that most previously published observations only include dust emission traced at millimeter wavelengths (1 and 3~mm bands), which may be optically thick and thus reduce the overall value of $\beta$. Since the young stars presented in the left panel of Figure~\ref{comparison_beta} boast some of the brightest disks observed in millimeter continuum emission, they represent a population of massive disks where high optical depth at millimeter wavelengths may be expected. Thus, including long-wavelength observations is critical for the calculation of $\beta(R)$, not only for the increased lever arm but also to avoid regions of high optical depth. We note that the comparison between previous results in the literature ($\beta_{\rm mm}(R)$) and our multiwavelength analysis ($\beta{\rm mm-cm}(R)$) should ideally be performed over a similar wavelength range to avoid potential biases between different wavelengths that trace different grain sizes, and could explain some of the differences between $\beta_{\mathrm{mm}}(R)$ and $\beta_{\mathrm{mm-cm}}(R)$ seen in Figure ~\ref{comparison_beta}.
Finally, we note that \cytau\ shows the steepest rise in $\beta$ with radius, indicative of a substantial change in dust properties over tens of AU in this disk, while the smoothest rise in the emissivity spectral index occurs for AS~209 in the inner disk and for \doardisk\ in the outer disk. However, these differences at not significant when the disks are compared at the $3\sigma$ level.

\section{Conclusions}
\label{conclusions}
We present multi-wavelength observations of the protoplanetary disks surrounding the young stars \cytau\ and \doardisk. 
Observations at 0.9, 1.3, 2.8, 8.0 and 9.8~mm spatially resolve the dust continuum emission from these disks, down to scales of tens of AU. 
From observations obtained at 5~cm, we quantify the amount of continuum emission whose origin is not thermal dust emission.
We estimate the level of contamination for \doardisk\ to be $<24$\% of the integrated emission at 8.0 and 9.8~mm, while for \cytau\ we establish an upper limit of contamination of $<4$\% of the total emission at 7.1~mm.
Although these levels of contamination may seem small, they need to be quantified when characterizing the dust emission spectrum.
The temperature and surface density profiles for both disks were constrained at all observed wavelengths shorter than 5~cm, and from this modeling, we find that a constant dust opacity with radius does not fit these observations. 
This result is supported by the observational fact that the normalized visibility profiles of the dust continuum emission differ for different wavelengths: 
cm-wave observations trace emission from a compact disk structure, while mm-wave observations trace a more extended disk structure. 

Since our modeling indicates that the dust continuum emission is optically thin over the spatial scales explored (and this is also warranted by the observed brightness temperature profiles in \S3.4), the observed emission spectrum will be directly proportional to the dust opacity spectrum, which allows us to infer
radial variations of the dust opacity spectral index, $\beta(R)$. We find that a constant value of $\beta$ is excluded by our observations
at more than $7\sigma$ in both disks. Close to the central protostar, for $R<50$~AU, the constraints on $\beta(R)$ are of high significance and
indicate a rapid change of the dust properties between the inner disk and the outer disk. In the outermost regions of the disk where the dust emission has lower SNR, we can only set an lower limit for the dust opacity spectral index
and we find that $\beta>1.0$ for \cytau\ at $R>80$~AU, while $\beta>1.2$ for \doardisk\ at $R>80$~AU.
We find that for the disks of AS~209, \cytau, and \doardisk, the profiles of $\beta(R)$ are steeper than for the available $\beta(R)$ constraints from the literature at the $1\sigma$ level \citep{2010Isella,2011Guilloteau,2013Trotta}, while these are all consistent at the $3\sigma$ level. The increased wavelength coverage presented here and by \citet{2012Perez} is critical for obtaining constraints of high significance on $\beta(R)$.

Assuming that the changes in the dust properties discussed above arise only from changes in the maximum particle size of the disk grain size distribution,
we derive radial variations of the maximum grain size, $a_{max}(R)$.
For the two different grain size distribution explored, $q=2.5$ and $q=3.5$, we find a gradient in $a_{max}$ with increasing grain size at smaller radial distance from the star, in both the \cytau\ and \doardisk\ disks. 
We compare the observational constraints in $a_{max}(R)$ with theoretical expectations of the maximum grain size when a population of grains grows and evolves while being limited by fragmentation or by radial drift. 
For an assumed disk turbulence of $\alpha_t=0.01$, the fragmentation barrier does not seem to be the limiting factor in the growth of these grains, as we observe millimeter and centimeter dust particles at distances from the star where the fragmentation barrier should have depleted these large particles. However, our observations seem to be consistent with a low-turbulence disk midplane with $\alpha_t\approx0.001$ in \cytau\ and \doardisk, as well as in AS~209 \citep{2012Perez}. If disks are generally low-turbulence, then fragmentation may play a lesser role in shaping the grain size distribution of protoplanetary disks. Future constraints on disk turbulence will be necessary to assess this process.
For the \cytau\ disk, the population of grains inferred from our observations is above the growth barrier imposed by radial drift and fragmentation, with a difference of of a factor of a few at $\gtrsim 100$~AU radius. We hypothesize that this difference could be readily explained if a region of higher gas pressure exists in the outer disk, and is keeping these millimeter-sized particles from drifting inwards. Future high resolution observations that trace millimeter dust grains, as well as the gas distribution, are needed. 
Observations with the Atacama Large sub-Millimeter Array (ALMA) will be extremely well suited for such investigations, as spatial resolutions matching the VLA observations at 7~mm (of $0.07''$ or better) are already possible \citep{2015arXiv150302649P}.
Finally, future observations of multiple disks at different stages of evolution (classical vs. transitional), as well as different stellar mass or disk masses, will provide valuable information related to the first steps toward planet formation.

\acknowledgments

We thank the referee for valuable comments. The National Radio Astronomy Observatory is a facility of the National Science Foundation operated under cooperative agreement by Associated Universities, Inc.
Ongoing CARMA development and operations are supported by the National Science Foundation under a cooperative agreement, and by the CARMA partner universities.
The SMA is a joint project between the Smithsonian Astrophysical Observatory and the Academia Sinica Institute of Astronomy and Astrophysics, funded by the Smithsonian Institution and Academia Sinica.
Part of this research was carried out at the Jet Propulsion Laboratory, Caltech, under a contract with the National Aeronautics and Space Administration.

{\it Facilities:} \facility{CARMA}, \facility{SMA}, \facility{VLA}.

\end{document}